\documentclass[a4paper,12pt]{article}

\usepackage[utf8]{inputenc}
\usepackage[english]{babel}
\usepackage{jheppub}
\usepackage{mathrsfs}
\usepackage{dsfont}
\usepackage{feynmp}
\usepackage{multirow}

\newcommand{\identity}{\mathds{1}}		
\newcommand{\I}{\mathcal{I}}			
\newcommand{\M}{\mathcal{M}}			
\newcommand{\N}{\mathcal{N}}			
\newcommand{\F}{\mathcal{F}}			
\renewcommand{\O}{\mathcal{O}} 		
\newcommand{\SO}{\text{SO}}
\newcommand{\timeorderingsymbol}{\text{T}}
\newcommand{\antitimeorderingsymbol}{\overline{\timeorderingsymbol}}
\newcommand{\timeordering}[1]{\timeorderingsymbol\big\{ #1 \big\}}
\newcommand{\antitimeordering}[1]{\antitimeorderingsymbol\big\{ #1 \big\}}
\newcommand{\vev}[1]{\big\langle 0 \big| #1 \big| 0 \big\rangle}
\newcommand{\ket}{\big| 0 \big\rangle}
\newcommand{\bra}{\big\langle 0 \big|}
\DeclareMathOperator{\re}{Re}
\DeclareMathOperator{\im}{Im}

\title{Momentum-space conformal blocks on the light cone}

\author{Marc Gillioz}

\affiliation{Theoretical Particle Physics Laboratory, Institute of Physics, 
EPFL, Lausanne, Switzerland}

\emailAdd{marc.gillioz@epfl.ch}

\date{\today}

\abstract{
We study the momentum-space 4-point correlation function of identical scalar operators in conformal field theory. Working specifically with null momenta, we show that its imaginary part admits an expansion in conformal blocks. The blocks are polynomials in the cosine of the scattering angle, with degree $\ell$ corresponding to the spin of the intermediate operator. The coefficients of these polynomials are obtained in a closed-form expression for arbitrary spacetime dimension $d > 2$. If the scaling dimension of the intermediate operator is large, the conformal block reduces to a Gegenbauer polynomial $\mathcal{C}_\ell^{(d-2)/2}$. If on the contrary the scaling dimension saturates the unitarity bound, the block is different Gegenbauer polynomial $\mathcal{C}_\ell^{(d-3)/2}$. These results are then used as an inversion formula to compute OPE coefficients in a free theory example.
}

\begin{document} 
\maketitle


\section{Introduction}
\label{sec:introduction}

The conformal bootstrap program in $d > 2$ spacetime dimensions exploits the crossing symmetry of 4-point correlation functions to constrain the space of conformal field theories (CFTs)~\cite{Rattazzi:2008pe}%
\footnote{See Ref.~\cite{Poland:2018epd} for a recent comprehensive review.}.
A key ingredient in this approach is the expansion of correlation functions into conformal blocks that are in one-to-one correspondence with conformal primaries in the operator product expansion (OPE)~\cite{Ferrara:1973vz, Ferrara:1974ny, Ferrara:1974nf}.
The computation of conformal blocks is notoriously difficult, and the result is often a complicated function of the position of the operators, for which there is not always a closed-form expression. 
This complexity has motivated the search for simpler formulations of the crossing equation, for instance using integration over the position of operators with various measures.
Some integration measures are specifically designed to make the crossing equations more tractable~\cite{Mazac:2016qev, Mazac:2018mdx}; others use insights from the AdS/CFT correspondence to represent the conformal blocks as Mellin integrals~\cite{Mack:2009mi, Penedones:2010ue, Dolan:2011dv, Fitzpatrick:2011ia, Paulos:2011ie, Fitzpatrick:2011hu}.
In this work, we would like to present a relatively simple method to derive conformal blocks based on the Fourier transform into momentum space.

The use of momentum-space techniques is standard in quantum field theory, following naturally from the necessity to implement translation symmetry in computations.
In conformal field theory, the 4-point functions are readily invariant under translations when expressed in terms of the conformal cross-ratios.
However, translation invariance is not trivial in the OPE: it is only recovered after summing the contributions of descendant operators. The direct computation of conformal blocks using the OPE is therefore rather cumbersome~\cite{Dolan:2000ut, Fortin:2016lmf, Fortin:2016dlj}, and in practice one usually prefers to use an approach based on solving a differential equation~\cite{Dolan:2003hv}. The momentum-space approach to conformal correlators make the direct computation of conformal blocks much simpler. Moreover, it allows to separate easily the Gaussian part of the correlation function given by the identity and double-trace operators from the rest of the theory.

But the use of momentum-space techniques in conformal field theory comes at a price, and there are two immediate difficulties that must be overcome. The first is of technical nature: while translations act trivially in momentum space, special conformal transformations are more involved, as their generators are second-order differential operators. The structure of 2- and 3-point correlation functions can still be derived from conformal Ward identities, but the procedure is not as simple as in position space~\cite{Bzowski:2012ih, Bzowski:2013sza, Bzowski:2015pba, Bzowski:2017poo, Bzowski:2018fql, Coriano:2013jba, Coriano:2018bbe, Kundu:2014gxa, Kundu:2015xta, Arkani-Hamed:2015bza}.
The second problem associated with momentum space has do with the very definition of the OPE. The standard argument based on radial quantization must be amended, and it does not seem that an OPE can be applied at all for time-ordered correlation functions.
There exist instead an approach that uses a crossing-symmetric basis of functions, following Polyakov's original bootstrap idea~\cite{Polyakov:1974gs}. Such a basis has been recently constructed using Witten diagrams and Mellin integral representations~\cite{Gopakumar:2016wkt, Gopakumar:2016cpb}, with an direct interpretation in momentum-space language~\cite{Isono:2018rrb}.
In this work, we will explore a different direction and show instead that both of the difficulties mentioned above are alleviated when one considers null momenta in Minkowski space, i.e.~momenta that lie on the (future or past) light cone. In this case the 3-point functions take a simpler form, and, more importantly, the imaginary part of the 4-point function can be expanded in conformal blocks~\cite{Gillioz:2016jnn, Gillioz:2018kwh}.

\begin{figure}
	\centering
	\begin{fmffile}{scattering}
		\begin{fmfgraph*}(250,80)
			\fmfset{arrow_len}{4mm}
			\fmfleft{l1,l2,l3}
			\fmfright{r1,r2,r3}
			\fmf{fermion,label=$\vec{p}_1$,l.s=left}{l2,v}
			\fmf{fermion,label=$\vec{p}_2$,l.s=left}{r2,v}
			\fmffreeze
			\fmf{fermion,label=$\vec{p}_3$,l.s=left}{v,r3}
			\fmf{fermion,label=$\vec{p}_4$,l.s=left}{v,l1}
			\fmfv{decor.shape=circle,decor.filled=shaded,decor.size=40,
			label=$\theta$, l.dist=32, l.angle=12}{v}
		\end{fmfgraph*}
	\end{fmffile}
	\caption{$2 \to 2$ scattering configuration in the center-of-mass frame,
	with incoming massless particles carrying momenta $p_1$ and $p_2$,
	and outgoing particles momenta $p_3$ and $p_4$.}
	\label{fig:scattering}
\end{figure}
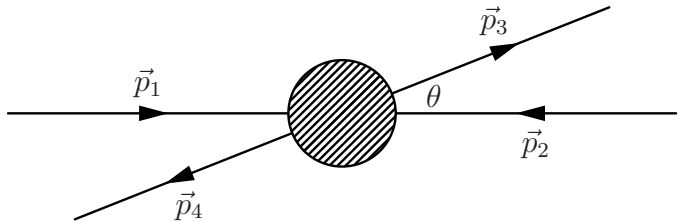

We will focus on the time-ordered correlation function of 4 identical scalar primary operators $\phi$,
\begin{equation}
	\vev{ \timeordering{ \phi(p_1) \phi(p_2) \phi(p_3) \phi(p_4) }}
	\equiv (2\pi)^d \delta^d(p_1 + p_2 + p_3 + p_4) i \M(p_1, p_2, p_3).
	\label{eq:4pt}
\end{equation}
This is equivalent to defining $\M$ as
\begin{equation}
	i \M(p_1, p_2, p_3)
	= \int d^dx_1 d^dx_2 d^dx_3
	e^{i(p_1 \cdot x_1 + p_2 \cdot x_2 + p_3 \cdot x_3)}
	\vev{ \timeordering{ \phi(x_1) \phi(x_2) \phi(x_3) \phi(0) }},
	\label{eq:4pt:alternative}
\end{equation}
where we have used translation invariance to move one point to the origin of the coordinate system.
By Lorentz invariance, $\M$ is a function of the ``masses'' $m_i^2 = -p_i^2$, and of the Mandelstam invariants%
\footnote{We are obviously working with the ``mostly plus'' convention for the metric in $d$ dimensions.}
\begin{equation}
	s = -(p_1 + p_2)^2,
	\qquad\quad
	t = -(p_1 + p_3)^2,
	\qquad\quad
	u = -(p_2 + p_3)^2,
	\label{eq:Mandelstam}
\end{equation}
For the reasons mentioned previously, we will restrict our analysis to null momenta ($p_i^2 = 0$), assuming for now that the Fourier transform exists in this limit. Two of the momenta must lie on the future light cone (we take these to be $p_1$ and $p_2$) and two on the past light cone ($p_3$ and $p_4$), so that $s > 0$ and $t, u \leq 0$. We will parameterize the Mandelstam invariants $t$ and $u$ as
\begin{equation}
	t = -\frac{s}{2} (1 - x),
	\qquad
	u = -\frac{s}{2} (1 + x),
	\qquad
	-1 \leq x \leq 1.
	\label{eq:xdef}
\end{equation}
so that the relation $s + t + u = 0$ is satisfied.
This kinematic configuration corresponds to a $2 \to 2$ scattering process in which $x = \cos\theta$, where $\theta$ is the scattering angle, as illustrated in Fig~\ref{fig:scattering}.

The existence of an OPE is related to the ability of defining states corresponding to a single operator insertion. Because of the time-ordering operator, a complete set of states cannot be inserted directly into the 4-point function~\eqref{eq:4pt}. Instead, we will use an OPE valid for the imaginary part of $\M$ only, which satisfies
\begin{equation}
	 (2\pi)^d \delta^d(p_1 + p_2 + p_3 + p_4) \, 2 \im \M(p_1, p_2, p_3)
	 = \vev{ \antitimeordering{ \phi(p_3) \phi(p_4) }
	\timeordering{ \phi(p_1) \phi(p_2) }}.
	\label{eq:4pt:imaginarypart}
\end{equation}
$\timeorderingsymbol$ ($\antitimeorderingsymbol$) indicate respectively the (anti-)time-ordered product of the operators.
The kinematic choice $s > 0$, $t \leq 0$ is implicit here.
By dimensional analysis, the imaginary part of $\M$ must satisfy%
\footnote{Note that if the quantity $2\Delta_\phi - 3d/2$ is an integer, there could be a scale anomaly in the 4-point function, which means a logarithmic dependence on $s$ in the real part of $\M$~\cite{Gillioz:2016jnn, Gillioz:2018kwh}. The imaginary part of $\M$ is nevertheless guaranteed to take the form of Eq.~\eqref{eq:G} in that case.}
\begin{equation}
	2 \im \M(p_1, p_2, p_3)
	= s^{2\Delta_\phi - 3d/2} \, G(x),
	\label{eq:G}
\end{equation}
where $\Delta_\phi$ is the scaling dimension of the operator $\phi$.
We will show that there exists a conformal block expansion for the dimensionless function $G(x)$, in the form
\begin{equation}
	G(x) = \sum_{\O \in \phi \times \phi} \lambda_{\phi\phi\O}^2 \, G_{\Delta, \ell}(x)
\end{equation}
where $\O$ indicates any primary operator that appears in the OPE of $\phi$, with scaling dimensions $\Delta$, spin $\ell$,
and OPE coefficient $\lambda_{\phi\phi\O}$.%
\footnote{Since there are only traceless symmetric tensors that enter the OPE of two scalar operators, the spin $\ell$ of the operator $\O$ is sufficient to characterize its Lorentz representation in any dimension $d$. Also, there is a single OPE coefficient associated with each operator $\O$.}
The result of our analysis is that the conformal block $G_{\Delta,\ell}(x)$ can be written as
\begin{equation}
	G_{\Delta,\ell}(x) = \N_{\Delta, \ell} \, g_{\Delta, \ell}(x)
	\label{eq:GNg}
\end{equation}
where $\N_{\Delta, \ell}$ is a normalization constant discussed below, and $g_{\Delta, \ell}(x)$ an even polynomial of degree $\ell$ in $x$,
\begin{equation}
	g_{\Delta,\ell}(x) = 
	\sum_{n = 0}^{\lfloor \ell / 2 \rfloor} \mathcal{X}_{\ell,n} \, x^{\ell-2n},
	\label{eq:g}
\end{equation}
with coefficients
\begin{eqnarray}
	\mathcal{X}_{\ell,n} & = & \frac{\ell! (2n)!}
	{2^{4n} (n!)^2 (\ell -2n)!
	\Big( \frac{3 - \Delta - \ell}{2} \Big)_n
	\Big( \frac{3 - \widetilde{\Delta} - \ell}{2} \Big)_n}
	\nonumber \\
	&& \qquad \times
	\, _3F_2\left( -n, 1 - \tfrac{\Delta + \ell}{2}, 1 - \tfrac{\widetilde{\Delta} + \ell}{2};
	\tfrac{1}{2} - n, 2 - \tfrac{d}{2} - \ell;
	1 \right).
	\label{eq:X}
\end{eqnarray}
written in terms of a generalized hypergeometric $_3F_2$ function%
\footnote{This hypergeometric function falls in the category of the so-called continuous dual Hahn polynomials~\cite{Gopakumar:2016cpb}. We thank Matthijs Hogervorst for pointing this out.}
and of the Pochhammer symbol $(a)_n = \Gamma(a + n) / \Gamma(a)$. $\widetilde{\Delta}$ is the shadow operator dimension defined by $\widetilde{\Delta} = d - \Delta$.
The coefficients $\mathcal{X}_{\ell,n}$ can alternatively be written as a finite sum presented in Eq.~\eqref{eq:X:sum}.
The polynomials $g_{\Delta,\ell}(x)$ are regular for every $\Delta$ satisfying the unitarity bound $\Delta \geq d - 2 + \ell$ (for $\ell > 0$), and they do not depend on the dimension $\Delta_\phi$ of the external operator.
They have several remarkable limits:
\begin{itemize}

\item
If the scaling dimension of the intermediate operator is large compared to its spin or to the dimension of spacetime, i.e.~$\Delta \gg \ell, d$, then $g_{\Delta, \ell}(x)$ can be written as
\begin{equation}
	g_{\infty, \ell}(x)
	= \frac{\ell !}{2^\ell \left( \frac{d-2}{2} \right)_\ell} \,
	\mathcal{C}_\ell^{(d-2)/2}(x),
	\label{eq:g:largeDelta}
\end{equation}
where $\mathcal{C}_\ell^{(d-2)/2}(x)$ is a Gegenbauer polynomial. This is the same Gegenbauer polynomial that appears in the position-space conformal block in the limit where two of the operators are close to each other~\cite{Dolan:2000ut, Costa:2011dw}.

\item
If the scaling dimension of the intermediate operator saturates the unitarity bound,
i.e.~if $\Delta = d - 2 + \ell$, then $g_{\Delta, \ell}(x)$ reduces to a different Gegenbauer polynomial, namely
\begin{equation}
	g_{d - 2 + \ell, \ell}(x)
	= \frac{\ell !}{2^\ell \left( \frac{d-3}{2} \right)_\ell} \,
	\mathcal{C}_\ell^{(d-3)/2}(x).
	\label{eq:g:unitaritybound}
\end{equation}
This case is strikingly analogous to Eq.~\eqref{eq:g:largeDelta} with the spacetime dimension lowered by one unit. This coincidence arises since a traceless symmetric tensor that satisfy a conservation condition effectively transform in the subgroup $\SO(d-1)$ of the Lorentz group $\SO(d-1,1)$, as explained in more detail in Section~\ref{sec:3point}.

\item
If the spacetime dimension is large while the scaling dimension of the operator remains at finite value above the unitarity bound ($\Delta \gtrsim d - 2 + \ell$), then only the leading power in $x$ remains in the polynomial:
\begin{equation}
	g_{\Delta, \ell}(x) \xrightarrow{d \to \infty} x^\ell.
	\label{eq:g:largeD}
\end{equation}

\end{itemize}

\begin{figure}
	\centering
	\includegraphics[width=0.49\linewidth]{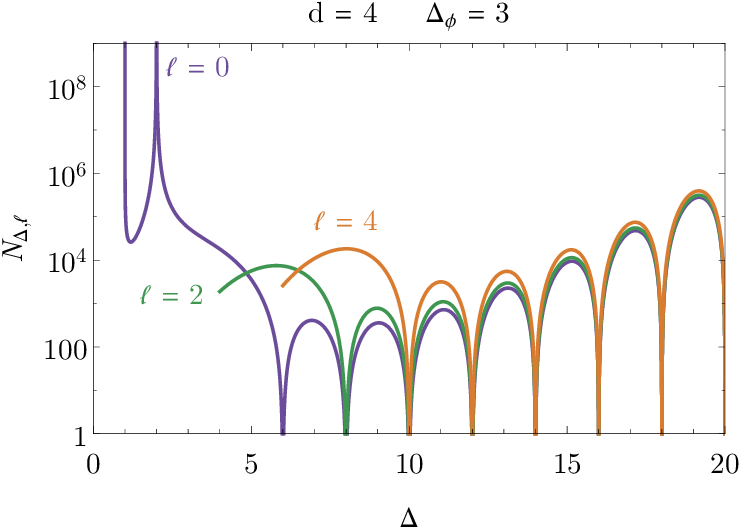}
	\includegraphics[width=0.49\linewidth]{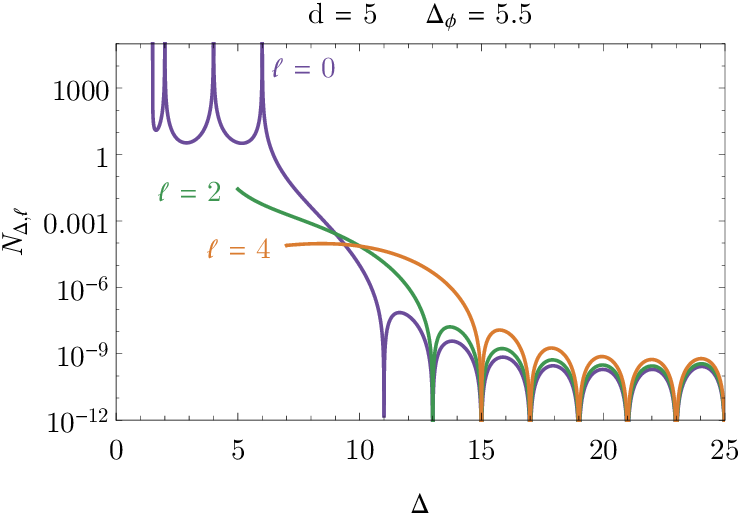}
	\caption{The coefficient $\N_{\Delta, \ell}$ of Eq.~\eqref{eq:N} as a function of $\Delta$
	and for $\ell = 0, 2$ and 4 in two different cases:
	the left panel corresponds to $d = 4$ spacetime dimensions
	with a scalar operator of scaling dimension $\Delta_\phi = 3$;
	the right panel to $d = 5$ and $\Delta_\phi = 5.5$.}
	\label{fig:N}
\end{figure}
The constant $\N_{\Delta, \ell}$ in Eq.~\eqref{eq:GNg} depends on the normalization of 2- and 3-point functions and is therefore related to the definition of the OPE coefficients. Working with standard conventions specified later, it is given by
\begin{equation}
	\N_{\Delta, \ell}
	= \frac{(4\pi)^{3d/2+1} \Gamma\left( \Delta_\phi - \frac{d}{2} \right)^4
	\Gamma\left( \Delta - \frac{d-2}{2} \right)
	\Gamma\left( \frac{\Delta + \ell}{2} - \Delta_\phi + \frac{d}{2} \right)^2
	\left( \Delta - 1 \right)_\ell
	\Gamma(\Delta + \ell)}
	{2^{4 \Delta_\phi + 2 \ell + 1}
	\Gamma\left( \frac{\Delta + \ell}{2} \right)^4
	\Gamma\left( \Delta_\phi - \frac{\Delta - \ell}{2} \right)^2
	\Gamma\left( \Delta_\phi + \frac{\Delta + \ell}{2} - \frac{d}{2} \right)^2
	\Gamma\left( \Delta_\phi + \frac{\Delta - \ell}{2} - d + 1 \right)^2}.
	\label{eq:N}
\end{equation}
Unlike $g_{\Delta, \ell}(x)$, this expression depends on $\Delta_\phi$ and is not always regular.
It is notably divergent if the dimension $\Delta_\phi$ of the external operator is $d/2$ or $(d-2)/2$. The latter case is trivial, as the operator $\phi$ must be a free field and $\M$ does not have an imaginary part; in the former case our approach is simply inconclusive, and we will therefore always assume $\Delta_\phi \neq \frac{d}{2}$ in this work.%
\footnote{When $\Delta_\phi = \frac{d}{2}$, the momentum-space 2-point function of $\phi$ must be renormalized. This can for instance be achieved shifting the scaling dimension of $\phi$ by an infinitesimal parameter, in which case the conformal block expansion that we derived should still be valid.}
Moreover, $\N_{\Delta, \ell}$ is divergent for intermediate operators with scaling dimension $\Delta = 2 \Delta_\phi - d - \ell - 2n$, with $n \in \mathbb{N}$. We will also assume that no such operator appear in the $\phi \times \phi$ OPE, which is a reasonable assumption if $\phi$ is taken to be a low-dimension operator of the CFT.
Besides its divergences, the coefficient $\N_{\Delta, \ell}$ has zeros whenever the intermediate operator has a dimension that matches one of the operators of the schematic double-trace form $\phi \partial^{2n} \phi$, i.e.~when $\Delta - \ell = 2 \Delta_\phi + 2n$. This property is in agreement with the fact that $\M$ must be trivial in a generalized free field theory where all correlators are Gaussian.
The zeros and singularities of $\N_{\Delta,\ell}$ are illustrated in two representative cases in Fig.~\ref{fig:N}. The figure also shows that $\N_{\Delta,\ell}$ becomes independent of the spin at large $\Delta$.

These are the results of our work, and their derivation is detailed in the remainder of the paper.
Section~\ref{sec:2point} is devoted to the study of 2-point functions in momentum space. It also contains a discussion of the OPE defined through the state/operator correspondence, and of the special role played by shadow operators.
In Section~\ref{sec:3point}, we derive an expression for the 3-point function of two scalar operators with a generic traceless symmetric spin-$\ell$ tensor, and compute the conformal blocks as products of pairs of 3-point functions.
Section~\ref{sec:freefields} provides a free theory example in which only higher-spin conserved current have to be considered.
The blocks are Gegenbauer polynomials in that case, and their orthogonality can be used to invert the OPE.
Some additional details regarding the computation of the 4-point function in terms of Feynman diagrams
are relegated to Appendix~\ref{sec:loopintegral}.
Finally, we conclude in Section~\ref{sec:conclusions} with a discussion of issues that were ignored before, such as the definiteness of the Fourier transform in the light-cone limit, related to the question of the OPE convergence in momentum space.


\section{Two-point functions and the momentum-space OPE}
\label{sec:2point}

In this section we discuss momentum-space operators and states, focusing on the simplest observables that are 2-point correlation functions. The goal is to derive a completeness relation that will later define the conformal block expansion. We begin with scalar operators, and then discuss traceless symmetric spin-$\ell$ tensors.

\subsection{The scalar two-point function}

In a Lorentzian theory, there are two distinct types of correlators depending on how light-like-separated points are treated. Specializing to scalar 2-point functions, one distinguishes the time-ordered correlator (or Feynman 2-point function)
\begin{equation}
	F_\Delta(x) \equiv \vev{ \timeordering{ \O(x) \O(0) }}
	= \frac{1}{[ x^2 + i \epsilon ]^\Delta}
	\label{eq:scalar2pt:position:Feynman}
\end{equation}
from the Wightman 2-point function, written without the time-ordering product,
\begin{equation}
	W_\Delta(x) \equiv \vev{ \O(x) \O(0) }
	= \frac{1}{[ -(x^0 - i \epsilon)^2 + (x^i)^2 ]^\Delta},
	\label{eq:scalar2pt:position:Wightman}
\end{equation}
where the infinitesimal $\epsilon > 0$ in both cases. These equations define the normalization of the primary operator $\O$.
The distinction between the two orderings is particularly important in momentum space, where we integrate over all of spacetime, including points at zero distance from each other.
The corresponding momentum-space correlators are
\begin{equation}
	F_\Delta(q) \equiv \int d^dx \, e^{-i q \cdot x} F_\Delta(x)
	= -i \frac{\pi^{d/2} \, \Gamma\left( \frac{d}{2} - \Delta \right)}
	{2^{2 \Delta - d} \, \Gamma\left( \Delta \right)}
	(q^2 - i \epsilon)^{\Delta - d/2}
	\label{eq:scalar2pt:Feynman}
\end{equation}
and
\begin{equation}
	W_\Delta(q) \equiv \int d^dx \, e^{-i q \cdot x} W_\Delta(x)
	= \Theta(q^0) \Theta(-q^2) \frac{\pi^{d/2+1}}
	{2^{2\Delta - d - 1} \Gamma\left( \Delta \right)
	\Gamma\left( \Delta - \frac{d-2}{2} \right)}
	(-q^2)^{\Delta - d/2},
	\label{eq:scalar2pt:Wightman}
\end{equation}
where $\Theta$ is the Heaviside step function ($\Theta(a) = 1$ for $a>0$, $\Theta(a) = 0$ otherwise).
The time-ordered 2-point function has support for all $q$, but it is divergent and needs renormalization whenever $\Delta = \frac{d}{2} + n$. On the contrary, the Wightman 2-point function only has support in the future momentum-space light cone, but it is well-defined and positive for all scaling dimensions $\Delta$ satisfying the unitarity bound. This positivity condition is necessary in any unitary theory since the Wightman 2-point function defines the norm of a state
\begin{equation}
	\big\langle \O(q') \big| \O(q) \big\rangle
	= (2\pi)^d \delta^d(q' + q) W_\Delta(q)
\end{equation}
where we have taken
\begin{equation}
	 \big| \O(q) \big\rangle \equiv \int d^dx \, e^{i q \cdot x}
	 \O(x^0 + i \epsilon, x^i) \ket,
	 \qquad
	 \big\langle \O(q) \big| = \big| \O(-q) \big\rangle^\dag.
	 \label{eq:state}
\end{equation}

An important property of CFTs in Minkowski space is that the set of states $\{ | \O(q) \rangle \}$ for all $q$ in the future light cone spans the full Verma module of the primary operator $\O$~\cite{Gillioz:2016jnn, Katz:2016hxp}. A simple way of seeing this is to insert this set of states in the position-space Wightman 2-point function \eqref{eq:scalar2pt:position:Wightman} and to verify that the full expression is recovered independently of the positions of the 2 points.%
\footnote{The exact form of the completeness relation $\identity \propto | \O(q) \rangle \langle \O(-q) |$ needed to perform this check will be given later in Eq.~\eqref{eq:completenessrelation}.}
Alternatively, it can be noted that the set of states~\eqref{eq:state} represents all linear combinations of a primary operator $\O$ inserted at the point $(x^0, x^i) = (i\epsilon, 0)$ and of its descendants. There exist a unitary evolution operator
\begin{equation}
	U(t) = e^{i H t},
	\qquad\qquad
	H = \epsilon^2 P_0 - K_0,
\end{equation}
that takes the constant time slice $x^0 = 0$ into spheres enclosing the point $(i \epsilon, 0)$ and its conjugate $(-i \epsilon, 0)$, with decreasing radius as $|t| \to \infty$.
Here $P_\mu$ and $K_\mu$ are respectively the generators of translations and special conformal transformations, and $H$ corresponds to the Hamiltonian of N-S quantization~\cite{Luscher:1974ez}.%
\footnote{Note that this construction works for any $\epsilon > 0$, not necessarily infinitesimal, but it is convenient to take $\epsilon \ll 1$ in the sense that it reproduces naturally the prescription of the Wightman function~\eqref{eq:scalar2pt:position:Wightman} in which Lorentz invariance is explicit.}
This construction defines a state/operator correspondence for Lorentzian CFTs, 
in exact analogy with radial quantization in Euclidean space.
States that carry spin can be constructed in an analogous way from operators with spin, and we will eventually be able to define a completeness relation that applies to arbitrary Wightman correlation functions.
In order to do so, we first need to determine the normalization of 2-point functions, i.e.~to compute the equivalent of Eq.~\eqref{eq:scalar2pt:Wightman} for operator with spin.

\subsection{The two-point function of traceless symmetric tensors}

The only operators with spin that need to be considered in this work are those that can appear in the OPE of two scalars, and they all belong to the traceless symmetric representations of $\SO(d-1,1)$.
The Wightman 2-point functions
\begin{equation}
	W_\Delta^{\mu_1 \ldots \mu_\ell,\nu_1 \ldots \nu_\ell}(x)
	\equiv \vev{ \O^{\mu_1 \ldots \mu_\ell}(x)
	\O^{\nu_1 \ldots \nu_\ell}(0) },
\end{equation}
are known to be
\begin{equation}
	W_\Delta^{\mu_1 \ldots \mu_\ell,\nu_1 \ldots \nu_\ell}(x)
	= W_\Delta(x) \, 
	\left[ \frac{1}{\ell!} \,
	\I^{\mu_1\nu_1}(x) \cdots \I^{\mu_\ell\nu_\ell}(x)
	+ \text{permutations} - \text{traces} \right]
	\label{eq:2pt:position}
\end{equation}
with
\begin{equation}
	\I^{\mu\nu}(x) = \eta^{\mu\nu} - 2 \, \frac{x^\mu x^\nu}{x^2},
\end{equation}
and where permutations and traces are understood to be among $\mu_i$ and $\nu_i$ indices separately.%
\footnote{We have chosen in this work to normalize all traceless symmetric operators such that their 2-point function is given by Eq.~\eqref{eq:2pt:position}. For conserved operators, such as the energy-momentum tensor or conserved currents associated with global symmetries, the standard normalization differs from our convention, as it is usually taken so that they satisfy canonical Ward identities. The conversion between these two normalizations can be simply achieved through a redefinition of OPE coefficients. 
}
There are several approaches to computing the Fourier transform of this 2-point function (see for instance Ref.~\cite{Isono:2018rrb}). We will present a method based on conformal Ward identities, which is convenient as it generalizes naturally to the computation of 3-point functions in Section~\ref{sec:3point}.
The idea is to consider the most general object that has the correct transformation properties under the Lorentz group, dilatation, and special conformal transformations in the form of the second-order differential equation%
\footnote{Invariance of the 2-point function under translations is ensured by the delta function imposing momentum conservation, which has been factored out of the definition of $W_\Delta^{\mu_1 \ldots \mu_\ell,\nu_1 \ldots \nu_\ell}(x)$.
}
\begin{equation}
	\left[ - q^\sigma \frac{\partial^2}{\partial q^\rho \partial q^\sigma}
	+ \frac{1}{2} q_\rho \frac{\partial^2}{\partial q_\sigma \partial q^\sigma}
	+ ( \Delta - d ) \frac{\partial}{\partial q^\rho}
	+ \frac{\partial}{\partial q_\sigma} \Sigma_{\rho\sigma}^{(\mu)} \right]
	W_\Delta^{\mu_1 \ldots \mu_\ell, \nu_1 \ldots \nu_\ell}(q) = 0,
	\label{eq:WardIdentity:2pt}
\end{equation}
where $\Sigma_{\rho\sigma}^{(\mu)}$ is the spin matrix acting on the indices $\mu_1 \ldots \mu_\ell$ only, not on $\nu_1 \ldots \nu_\ell$.
The unique solution to this problem, up to an overall normalization constant, is
\begin{eqnarray}
	W_\Delta^{\mu_1 \ldots \mu_\ell,\nu_1 \ldots \nu_\ell}(q)
	& = &  C_\O \, \Theta(q^0) \Theta(-q^2) (-q^2)^{\Delta - d/2}
	\sum_{n=0}^\ell \frac{2^n \ell!}{n! (\ell - n)!}
	\frac{\left( \frac{d}{2} - \Delta \right)_n}
	{\left( 2 - \Delta - \ell \right)_n}
	\label{eq:2pt:momentum}
	\\
	&& \quad \times \left[
	\frac{1}{\ell!}
	\frac{q^{\mu_1} q^{\nu_1} \cdots q^{\mu_n} q^{\nu_n}}{(-q^2)^n} \,
	\eta^{\mu_{n+1}\nu_{n+1}} \cdots \eta^{\mu_\ell\nu_\ell}
	+ \text{permutations} - \text{traces}
	\right].
	\nonumber 
\end{eqnarray}
The constant $C_\O$ can then be determined by contracting the indices of both operators: One the one hand, from Eq.~\eqref{eq:2pt:position},
\begin{equation}
	\eta_{\mu_1\nu_1} \cdots \eta_{\mu_\ell\nu_\ell}
	W_\Delta^{\mu_1 \ldots \mu_\ell,\nu_1 \ldots \nu_\ell}(x)
	= \frac{(d-2)_\ell}{\ell!} \,
	 W_\Delta(x).
\end{equation}
On the other hand, from  Eq.~\eqref{eq:2pt:momentum},
\begin{equation}
	\eta_{\mu_1\nu_1} \cdots \eta_{\mu_\ell\nu_\ell}
	W_\Delta^{\mu_1 \ldots \mu_\ell,\nu_1 \ldots \nu_\ell}(q)
	= C_\O \, \Theta(q^0) \Theta(-q^2) 
	\frac{\left( d - 2 \right)_\ell (\Delta + \ell - 1)}
	{\ell ! (\Delta - 1)} \,
	(-q^2)^{\Delta - d/2}.
\end{equation}
Making use of the Fourier transform~\eqref{eq:scalar2pt:Wightman} of the scalar 2-point function, one can deduce that
\begin{equation}
	C_\O
	= \frac{\pi^{d/2+1}}
	{2^{2 \Delta - d - 1} (\Delta + \ell - 1)
	\Gamma\left( \Delta - 1 \right)
	\Gamma\left( \Delta - \frac{d-2}{2} \right)},
	\label{eq:CO}
\end{equation} 
provided that $q^0 > 0$ and that $q^2 < 0$.
We have thus obtained an expression valid for any traceless symmetric tensor. The momentum-space 2-point function of a vector field is for instance
\begin{equation}
	W_\Delta^{\mu,\nu}(q)
	= \Theta(q^0) \Theta(-q^2) (-q^2)^{\Delta - d/2} \frac{\pi^{d/2+1} (\Delta - 1)}
	{2^{2 \Delta_ - d - 1}
	\Gamma\left( \Delta + 1 \right)
	\Gamma\left( \Delta - \frac{d-2}{2} \right)}
	\left[ \eta^{\mu\nu} + \frac{d - 2 \Delta}{\Delta - 1} \frac{q^\mu q^\nu}{q^2} \right],
\end{equation}
and it can be verified that the conservation condition $q_\mu W_\Delta^{\mu,\nu}(q)$ is automatically satisfied when $\Delta = d - 1$. Note that there is no simple factorization of the tensor structure as  in the position-space expression~\eqref{eq:2pt:position}.

\subsection{Completeness relation and shadow operators}

The existence of a completeness relation can be inferred from the state/operator correspondence discussed above. The states~\eqref{eq:state} and their generalization $\big| \O^{\mu_1 \ldots \mu_\ell}(k) \big\rangle$ for operators with spin readily satisfy orthogonality properties, both for different primary operators and for unequal momenta. We have therefore
\begin{equation}
	\identity = |0 \rangle \langle 0 | +  \sum_\O
	\int\limits_{\substack{q^0 >0 \\ q^2 < 0}} \frac{d^dq}{(2\pi)^d}
	\Pi^\Delta_{\mu_1 \ldots \mu_\ell, \nu_1 \ldots \nu_\ell}(q)
	\big| \O^{\mu_1 \ldots \mu_\ell}(q) \big\rangle
	\big\langle \O^{\nu_1 \ldots \nu_\ell}(-q) \big|,
	\label{eq:completenessrelation}
\end{equation}
where the sum is over all primary operators $\O \neq \identity$, and the tensors $\Pi(q)$ take into account the normalization of operators. They must be chosen such that
\begin{equation}
	W_\Delta^{\mu_1 \ldots \mu_\ell,\rho_1 \ldots \rho_\ell}(q)	
	\Pi^\Delta_{\rho_1 \ldots \rho_\ell, \sigma_1 \ldots \sigma_\ell}(q)
	W_\Delta^{\sigma_1 \ldots \sigma_\ell,\nu_1 \ldots \nu_\ell}(q)
	= W_\Delta^{\mu_1 \ldots \mu_\ell,\nu_1 \ldots \nu_\ell}(q).
\end{equation}
In our case, since we only consider scalar or traceless symmetric tensors, the $\Pi(q)$ can be determined from Eq.~\eqref{eq:2pt:momentum}. The solution is unique if we require that they transform under irreducible representations of the Lorentz group, i.e.~that they are traceless and symmetric in both sets of indices. By construction, we find
\begin{eqnarray}
	\Pi^\Delta_{\mu_1 \ldots \mu_\ell, \nu_1 \ldots \nu_\ell}(q)
	& = & \frac{(-q^2)^{d/2 - \Delta}}{C_\O}
	\sum_{n=0}^\ell \frac{2^n \ell!}{n! (\ell - n)!}
	\frac{\left( \Delta - \frac{d}{2} \right)_n}
	{\left( \Delta - \ell - d + 2 \right)_n}
	\label{eq:Pi}
	\\
	&& \quad \times \left[ \frac{1}{\ell!}
	\frac{q_{\mu_1} q_{\nu_1} \cdots q_{\mu_n} q_{\nu_n}}{(-q^2)^n} \,
	\eta_{\mu_{n+1}\nu_{n+1}} \cdots \eta_{\mu_\ell\nu_\ell}
	+ \text{permutations} - \text{traces}
	\right].
	\nonumber 
\end{eqnarray}
This tensor is singular when the dimension $\Delta$ saturates the unitarity bound, i.e.~when $\Delta = d - 2 + \ell$. In that case, however, the operator $\O^{\mu_1 \ldots \mu_\ell}$ is a conserved tensor and the corresponding state satisfies $q_{\mu_1} \big| \O^{\mu_1 \ldots \mu_\ell}(q) \big\rangle = 0$,
so that we can take 
\begin{equation}
	\Pi^{d - 2 + \ell}_{\mu_1 \ldots \mu_\ell, \nu_1 \ldots \nu_\ell}(q)
	= \frac{(-q^2)^{-d/2 + 2 - \ell}}{C_\O}
	\left[ \frac{1}{\ell!} \, \eta_{\mu_1\nu_1} \cdots \eta_{\mu_\ell\nu_\ell}
	+ \text{permutations} - \text{traces} \right].
\end{equation}
Alternatively, one can proceed with the expression~\eqref{eq:Pi} for generic $\Delta$ and take the limit $\Delta \to d - 2 + \ell$ at the end, as we will see that this gives finite results.

Comparing the tensor~\eqref{eq:Pi} with the 2-point function~\eqref{eq:2pt:momentum}, one can see that the former is obtained replacing $\Delta$ with $\widetilde{\Delta} = d - \Delta$ in the latter, up to the overall normalization coefficient, i.e.~$\Pi^\Delta_{\mu_1 \ldots \mu_\ell, \nu_1 \ldots \nu_\ell}(q) \propto W_{d-\Delta}^{\mu_1 \ldots \mu_\ell, \nu_1 \ldots \nu_\ell}(q)$.
This is not an accident but follows from the existence of a non-local ``shadow'' operator $\widetilde{\O}$ that has the same transformation properties as $\O$ under the conformal group, but with scaling dimension $\widetilde{\Delta} = d - \Delta$~\cite{Ferrara:1972xe, Ferrara:1972uq, Ferrara:1972ay, Ferrara:1973vz, SimmonsDuffin:2012uy}.
If we define ``shadow states''  by
\begin{equation}
	\big| \widetilde{\O}_{\mu_1 \ldots \mu_\ell}(q) \big\rangle
	\equiv
	\Pi^\Delta_{\mu_1 \ldots \mu_\ell, \nu_1 \ldots \nu_\ell}(q)
	\big| \O^{\nu_1 \ldots \nu_\ell}(q) \big\rangle,
	\label{eq:shadowstate}
\end{equation}
then the completeness relation can be expressed in the very simple form
\begin{equation}
	\identity = |0 \rangle \langle 0 | +  \sum_\O
	\int\limits_{\substack{q^0 >0 \\ q^2 < 0}} \frac{d^dq}{(2\pi)^d}
	\big| \widetilde{\O}_{\mu_1 \ldots \mu_\ell}(q) \big\rangle
	\big\langle \O^{\mu_1 \ldots \mu_\ell}(-q) \big|.
	\label{eq:completenessrelation:shadow}
\end{equation}
This alternative formulation is more than just a rewriting of the completeness relation \eqref{eq:completenessrelation}: since correlation functions $\langle \cdots \widetilde{\O} \rangle$ involving the shadow operator have similar transformation properties under the conformal group as the functions $\langle \cdots \O \rangle$, we will be able to determine the former directly in terms of the latter in Section~\ref{sec:3point}.

This concludes the derivation of a completeness relation that can be used to write an OPE for any Wightman correlation function.

\subsection{An OPE for the time-ordered 4-point function}

The crossing-symmetric 4-point function \eqref{eq:4pt} is not a Wightman correlation function, and therefore the completeness relation~\eqref{eq:completenessrelation:shadow} cannot be directly used to generate a conformal block expansion.
In position space, a time-ordered product can be expressed as a sum of Wightman functions multiplied with Heaviside step functions enforcing the chronological ordering. But this does not translate into a 
sum of momentum-space Wightman functions upon Fourier transform.
Instead, we make use of the combinatoric identity
\begin{eqnarray}
	\vev{ \timeordering{ \phi(x_1) \phi(x_2) \phi(x_3) \phi(x_4) }}
	+ \vev{ \antitimeordering{ \phi(x_1) \phi(x_2) \phi(x_3) \phi(x_4) }}
	&& \nonumber \\
	+ \Big( \vev{ \antitimeordering{ \phi(x_1) \phi(x_2) }
	\timeordering{ \phi(x_3) \phi(x_4) }}
	+ \text{permutations} \Big)
	&& \nonumber \\
	- \Big( \vev{ \phi(x_1) \timeordering{ \phi(x_2) \phi(x_3) \phi(x_4) }}
	+ \text{permutations} \Big)
	&& \nonumber \\
	- \Big( \vev{ \antitimeordering{ \phi(x_2) \phi(x_3) \phi(x_4) } \phi(x_1) }
	+ \text{permutations} \Big)
	& = & 0,
	\label{eq:timeorderingidentity}
\end{eqnarray}
where the permutations are among the $x_i$, and $\antitimeorderingsymbol$ denotes the anti-time-ordering operator. For real scalar operators, $\antitimeorderingsymbol$ corresponds to the Hermitian conjugate of the time-ordered product $\timeorderingsymbol$.
The Fourier transform of this equation relates the real part of the 4-point function (the first line) to a set of correlators that are of mixed Feynman/Wightman type. The correlators of the third and fourth lines vanish in the limit $p_i^2 \to 0$ if we approach it from the Euclidean side ($-p_i^2 < 0$), because in that case $\phi(p_i) \ket = 0$ and $\bra \phi(p_i)  = 0$.
The completeness relation \eqref{eq:completenessrelation:shadow} can then be used to evaluate each of the 6 terms of the second line as products of 3-point functions. Only one of them is non-zero, since $\timeordering{ \phi(p_i) \phi(p_j) } | 0 \rangle = 0$ if the combined momentum $p_i + p_j$ does not lie in the future light cone.
The only remaining term gives the equality
\begin{equation}
	2 \re \vev{ \timeordering{ \phi(p_1) \phi(p_2) \phi(p_3) \phi(p_4) }}
	= - \vev{ \antitimeordering{ \phi(p_3) \phi(p_4) }
	\timeordering{ \phi(p_1) \phi(p_2) }},
\end{equation}
which is equivalent to Eq.~\eqref{eq:4pt:imaginarypart} when written in terms of $\M(p_1, p_2, p_3)$.
Using the completeness relation on the right-hand side of this equation and performing the trivial integral over the exchange momentum $q$, one obtains finally
\begin{eqnarray}
	2 \im \M(p_1, p_2, p_3)
	& = & \sum_\O 
	\int d^dx_3 d^dx_4 \, e^{i (p_3 \cdot x_3 + p_4 \cdot x_4)}
	\vev{ \antitimeordering{ \phi(x_3) \phi(x_4) } \widetilde{\O}_{\mu_1 \ldots \mu_\ell}(0) }
	\nonumber \\
	&& \quad \times 
	\int d^dx_1 d^dx_2 \, e^{i (p_1 \cdot x_1 + p_2 \cdot x_2)}
	\vev{ \O^{\mu_1 \ldots \mu_\ell}(0) \timeordering{ \phi(x_1) \phi(x_2) } }.
	\nonumber \\
	\label{eq:4pt:factorization}
\end{eqnarray}
This is the essential equality that defines the conformal block expansion for the imaginary part of $\M$. It only involves 3-point functions in which the momentum-conserving delta functions have been factored out. Evaluating these functions is the subject of the next section.


\section{Three-point functions and conformal blocks}
\label{sec:3point}

In this section we describe the derivation of the 3-point functions of two scalars and one traceless symmetric spin-$\ell$ operator and the computation of their product as in Eq.~\eqref{eq:4pt:factorization}.
As for the 2-point function, we begin with the scalar case and later discuss the implementation of operators with spin.

\subsection{Scalar three-point function}

\begin{figure}
	\centering
	\begin{fmffile}{trianglediagram}
		\fmfframe(10,10)(10,10){
		\begin{fmfgraph*}(180,60)
			\fmfset{arrow_len}{3mm}
			\fmfleft{x2,x1}
			\fmfright{y2,x3,y1}
			\fmfv{label=$\phi$,l.a=180}{x1,x2}
			\fmfv{label=$\O$,l.a=0}{x3}
			\fmf{scalar,label=$p_1$,l.s=left}{x1,v1}
			\fmf{scalar,label=$p_2$,l.s=right}{x2,v2}
			\fmf{phantom,tension=0.6}{v1,y1}
			\fmf{phantom,tension=0.6}{v2,y2}
			\fmffreeze			
			\fmf{fermion,label=$k$,l.s=left}{v2,v1}
			\fmf{heavy,label=$p_1+k$,l.s=left}{v1,v3}
			\fmf{heavy,label=$p_2-k$,l.s=right}{v2,v3}
			\fmf{scalar,label=$p_1 + p_2$,tension=1.5}{v3,x3}
			\fmfdot{v1,v2,v3}
		\end{fmfgraph*}
		}
	\end{fmffile}
	\caption{Feynman diagram representation of Eq.~\eqref{eq:triangleintegral},
	in which the time-ordered 2-point function is indicated with a single solid line, and the 
	Wightman 2-point functions with double lines.}
	\label{fig:trianglediagram}
\end{figure}
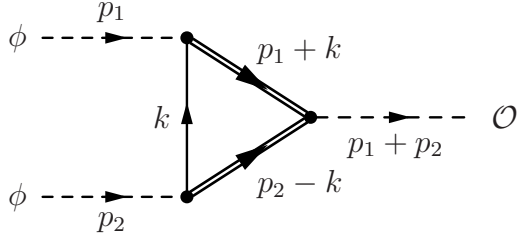

We denote the momentum-space 3-point function of scalar operators in which two of the operators are time-ordered by
\begin{equation}
	i \lambda_{\phi\phi\O} V_\Delta(p_1, p_2) =
	\int d^dx_1 d^dx_2 e^{i (p_1 \cdot x_1 + p_2 \cdot x_2)}
	\vev{ \O(0) \timeordering{ \phi(x_1) \phi(x_2) }}.
	\label{eq:V:scalar}
\end{equation}
The OPE coefficient $\lambda_{\phi\phi\O}$ has been taken out of the definition so that $V_\Delta(p_1, p_2)$ is a function of the scaling dimensions $\Delta \equiv \Delta_\O$ and $\Delta_\phi$, and of the momenta only.
The position-space 3-point function is given by
\begin{eqnarray}
	&& \vev{ \O(0) \timeordering{ \phi(x_1) \phi(x_2) } } 
	\\
	&& \quad
	=  \frac{\lambda_{\phi\phi\O} }
	{\left[ -(x_1^0 + i \epsilon)^2 + (x_1^i)^2 \right]^{\Delta/2}
	\left[ -(x_2^0 + i \epsilon)^2 + (x_2^i)^2 \right]^{\Delta/2}
	\left[ (x_1 - x_2)^2 + i \epsilon \right]^{\Delta_\phi - \Delta/2}}.
	\nonumber
\end{eqnarray}
For space-like separated points, this is the ordinary CFT 3-point function for scalar primary operators. In the general case, the $i \epsilon$ prescriptions ensure the correct ordering of operators. Using the translation invariance of the 3-point function, one can rewrite Eq.~\eqref{eq:V:scalar} in the form of a momentum integral over a product of 2-point functions as
\begin{equation}
	i V_\Delta(p_1, p_2) = \int \frac{d^dk}{(2\pi)^d}
	F_{\Delta_\phi - \Delta/2}(k)
	W_{\Delta/2}(p_1 + k)
	W_{\Delta/2}(p_2 - k),
	\label{eq:triangleintegral}
\end{equation}
where $F_\alpha$ and $W_\alpha$ are the time-ordered and Wightman 2-point functions given in Eqs.~\eqref{eq:scalar2pt:Feynman} and \eqref{eq:scalar2pt:Wightman} respectively, for fictitious scalar operators with scaling dimension $\alpha$.
This integral can be represented by the Feynman diagram in Fig.~\ref{fig:trianglediagram}.
It has been computed in Ref.~\cite{Gillioz:2016jnn} for null momenta $p_1^2 = p_2^2 = 0$, and the result can be written in terms of the invariant $s = -2p_1 \cdot p_2$ as
\begin{equation}
	V_\Delta(p_1, p_2) 
	= \frac{ 2^{2d - 2\Delta_\phi - \Delta + 1} \pi^{d+1}
	\Gamma\left( \Delta_\phi - \frac{d}{2} \right)^2
	\Gamma\left( \frac{\Delta}{2} - \Delta_\phi + \frac{d}{2} \right)}
	{\Gamma\left( \frac{\Delta}{2} \right)^2
	\Gamma\left( \Delta_\phi - \frac{\Delta}{2} \right)
	\Gamma\left( \Delta_\phi + \frac{\Delta}{2} - \frac{d}{2} \right)
	\Gamma\left( \Delta_\phi + \frac{\Delta}{2} - d + 1 \right)} \,
	s^{\Delta_\phi + \Delta/2 - d}.
	\label{eq:V}
\end{equation}
Combining this result with the OPE expansion~\eqref{eq:4pt:factorization} for the imaginary part of $\M$, one can immediately compute the conformal block for an intermediate scalar operator to be 
\begin{equation}
	G_{\Delta,0}(x) = \frac{V_\Delta(p_1, p_2)^2}
	{s^{2\Delta_\phi + \Delta - 2d} C_\O},
\end{equation}
where $C_\O$ is given in Eq.~\eqref{eq:CO}. It is straightforward to verify that this expression coincides with Eq.~\eqref{eq:GNg} at $\ell = 0$.

\subsection{Three-point function for intermediate operators with spin}

The generalization of Eq.~\eqref{eq:V:scalar} for an operator $\O$ with spin will be denoted by
\begin{equation}
	i  \lambda_{\phi\phi\O} V_\Delta^{\mu_1 \ldots \mu_\ell}(p_1, p_2)
	= \int d^dx_1 d^dx_2 e^{i (p_1 \cdot x_1 + p_2 \cdot x_2)}
	\vev{ \O^{\mu_1 \ldots \mu_\ell}(0) \timeordering{ \phi(x_1) \phi(x_2) }},
	\label{eq:V:spinning}
\end{equation}
where the position-space 3-point function is given by~\cite{Osborn:1993cr, Costa:2011mg}
\begin{eqnarray}
	&& \vev{ \O^{\mu_1 \ldots \mu_\ell}(0) \timeordering{ \phi(x_1) \phi(x_2) } } 
	\label{eq:3pt:position}
	\\
	&&
	=  \frac{\lambda_{\phi\phi\O} 
	\left( R^{\mu_1} \cdots R^{\mu_\ell} - \text{traces} \right)}
	{\left[ -(x_1^0 + i \epsilon)^2 + (x_1^i)^2 \right]^{(\Delta - \ell)/2}
	\left[ -(x_2^0 + i \epsilon)^2 + (x_2^i)^2 \right]^{(\Delta - \ell)/2}
	\left[ (x_1 - x_2)^2 + i \epsilon \right]^{\Delta_\phi - (\Delta - \ell)/2}}.
	\nonumber
\end{eqnarray}
with
\begin{equation}
	R^\mu = \frac{x_1^\mu}{x_1^2} - \frac{x_2^\mu}{x_2^2}.
\end{equation}
There are various ways of computing the tensor $V_\Delta^{\mu_1 \ldots \mu_\ell}$.
One of them consists in expressing it as a differential operator acting on the scalar 3-point function~\cite{Isono:2018rrb}.
We will use instead conformal Ward identities in momentum space, following a strategy developed in Refs.~\cite{Bzowski:2012ih, Bzowski:2013sza, Bzowski:2015pba, Bzowski:2017poo, Bzowski:2018fql}.
Requiring Lorentz and scale invariance is simple enough, and we can parameterize the 3-point function as%
\footnote{Again, we assume that there is no scale anomaly in the 3-point function. }
\begin{eqnarray}
	V_\Delta^{\mu_1 \ldots \mu_\ell}(p_1,p_2)
	& = & s^{(2 \Delta_\phi + \Delta - 2d - \ell)/2}
	\sum_{n=0}^\ell \frac{(-1)^n \ell !}{n! (\ell - n)!}
	\F_{\ell, n}\left( \frac{-p_1^2}{s}, \frac{-p_2^2}{s} \right) 
	\nonumber \\
	&& \qquad \times
	\Big[ \frac{1}{\ell!} \, p_1^{\mu_1} \cdots p_1^{\mu_n} p_2^{\mu_{n+1}} \cdots p_2^{\mu_\ell}
	+ \text{permutations} - \text{traces} \Big],
\end{eqnarray}
where the $\F_{\ell, n}$ are functions of two dimensionless variables, and they are unknown at this stage.
Invariance of this expression under special conformal transformation gives additional constraints among the $\F_{\ell, n}$.
It is important to remark that since special conformal transformations do not preserve the light cone condition $p_i^2 = 0$, we must consider the general kinematics at arbitrary $p_i^2$ in order to derive these constraints. The Ward identity associated with special conformal transformations generated by $K_\rho$ is given by
\begin{equation}
	\sum_{i=1}^2 \left[ - p_i^\sigma \frac{\partial^2}{\partial p_i^\rho \partial p_i^\sigma}
	+ \frac{1}{2} p_{i\rho} \frac{\partial^2}{\partial p_{i\sigma} \partial p_i^\sigma}
	+ ( \Delta_\phi - d ) \frac{\partial}{\partial p_i^\rho} \right]
	V_\Delta^{\mu_1 \ldots \mu_\ell}(p_1, p_2) = 0.
	\label{eq:WardIdentity:3pt}
\end{equation}
Note that the differential operator acting on $V$ does not depend on the spin of the operator: this is because we have used translation invariance to place $\O$ at the origin of coordinate space, where $\left[ K_\rho, \O^{\mu_1 \ldots \mu_\ell}(0) \right] = 0$.
In general, the constraints among the functions $\F_{\ell,n}$ take the form of second order differential equations. In the limit $p_i^2 \to 0$, they reduce however to first order differential equations, because the second-order derivatives part of Eq.~\eqref{eq:WardIdentity:3pt} coincides with the Todorov operator that preserves the condition $p_i^2 = 0$~\cite{Dobrev:1975ru, Dymarsky:2017yzx}.
We will assume that the $\F_{\ell,n}$ are well-defined in that limit.
In order to simplify the problem further, the Ward identity can be split into components corresponding to special conformal transformations along $p_1$, $p_2$, and along orthogonal directions.
For instance, contracting Eq.~\eqref{eq:WardIdentity:3pt} with $p_2^\rho$, one obtains the condition
\begin{eqnarray}
	\left( \Delta_\phi - \tfrac{d}{2} - 1 \right) \partial_z \F_{\ell,n}(z,0) \Big|_{z = 0}
	& = & \left( \tfrac{\Delta - \ell}{2} + \Delta_\phi - d \right)
	\left( \tfrac{\Delta + \ell}{2} - \Delta_\phi + \tfrac{d}{2} - n \right) \F_{\ell,n}(0,0) 
	\nonumber \\
	&& + ( \ell - n ) \left( \tfrac{\Delta - \ell}{2} + \ell - n - 1 \right) \F_{\ell,n}(0,0)
	\nonumber \\
	&& + ( \ell - n ) ( \Delta_\phi - d - n ) \F_{\ell, n+1}(0,0),
\end{eqnarray}
which relates first derivatives of the $\F_{\ell,n}$ with their values at $p_i^2 = 0$. A similar equation is obtained when contracting Eq.~\eqref{eq:WardIdentity:3pt} with $p_1^\rho$, in this case involving derivatives of the $\F_{\ell,n}$ with respect to their second variable.
These equations always have a solution for generic $\Delta_\phi$, but they do not constrain the $\F_{\ell,n}$ at $p_i^2 = 0$.
The Ward identity in the orthogonal direction is more interesting: defining a vector $p_\perp$ such that $p_1 \cdot p_\perp = p_2 \cdot p_\perp = 0$ (which is always possible in $d > 2$) and  contracting it with Eq.~\eqref{eq:WardIdentity:3pt} leads to the condition
\begin{equation}
	\left( \tfrac{\Delta - \ell}{2} + n \right) \F_{\ell, n+1}(0,0)
	= \left( \tfrac{\Delta + \ell}{2} - n - 1 \right) \F_{\ell, n}(0,0).
\end{equation}
This recursion relation determines the 3-point function completely up to an overall normalization constant $C_{\phi\phi\O}$, and we get
\begin{eqnarray}
	V_\Delta^{\mu_1 \ldots \mu_\ell}(p_1,p_2) 
	& = & C_{\phi\phi\O} \,
	s^{\Delta_\phi - d + (\Delta - \ell)/2}
	\sum_{n=0}^\ell \frac{(-1)^{\ell - n} \ell !}{n! (\ell - n)!}
	\frac{2^\ell \left( \frac{\Delta + \ell}{2} - n \right)_n
	\left( \frac{\Delta - \ell}{2} + n \right)_{\ell - n}}{\left( \Delta - 1 \right)_\ell}
	\nonumber \\
	&& \qquad \times
	\left[ \frac{1}{\ell!} \, p_1^{\mu_1} \cdots p_1^{\mu_n}
	p_2^{\mu_{n+1}} \cdots p_2^{\mu_\ell}
	+ \text{permutations} - \text{traces} \right].
	\label{eq:3pt:p1p2}
\end{eqnarray}
It will turn out to be more convenient to express this 3-point function in terms of the sum and difference of the momenta $p_1$ and $p_2$, defining $q = p_1 + p_2$ and $r = p_1 - p_2$, for which
\begin{eqnarray}
	V_\Delta^{\mu_1 \ldots \mu_\ell}(q,r) 
	& = & C_{\phi\phi\O} \,
	s^{\Delta_\phi - d + (\Delta - \ell)/2}
	\sum_{n=0}^{\ell/2} \frac{\ell !}{n! (\ell - 2n)!}
	\frac{1}{2^{2n} \left( \frac{3 - \Delta - \ell}{2} \right)_n}
	\label{eq:3pt}
	\\
	&& \qquad \times
	\left[ \frac{1}{\ell!} \, q^{\mu_1} \cdots q^{\mu_{2n}}
	r^{\mu_{2n+1}} \cdots r^{\mu_\ell}
	+ \text{permutations} - \text{traces} \right].
	\nonumber
\end{eqnarray}
It is interesting to note that the tensor structure of the 3-point function does not depend on $\Delta_\phi$. This is a known feature of 3-point function involving two identical scalar operators.

In order to determine $C_{\phi\phi\O}$, we consider the scalar integral obtained contracting the symmetric tensor $p_1^{\mu_1} \cdots p_1^{\mu_\ell}$ with the 3-point function. On the one hand, using Eq.~\eqref{eq:3pt:p1p2}, we find
\begin{equation}
	p_{1\mu_1} \cdots p_{1\mu_\ell} V_\Delta^{\mu_1 \ldots \mu_\ell}(p_1,p_2) 
	= C_{\phi\phi\O} \frac{\left( \frac{\Delta - \ell}{2} \right)_\ell}
	{\left( \Delta - 1 \right)_\ell} \, s^{\Delta_\phi + (\Delta + \ell)/2 - d}.
	\label{eq:3pt:scalarcontraction:1}
\end{equation}
On the other hand, from the definition~\eqref{eq:V:spinning},
\begin{eqnarray}
	&& i \lambda_{\phi\phi\O} 
	p_{1\mu_1} \cdots p_{1\mu_\ell} V_\Delta^{\mu_1 \ldots \mu_\ell}(p_1,p_2) 
	\\
	&& \quad
	= i^{\ell} \int d^dx_1 d^dx_2 \, e^{i (p_1 \cdot x_1 + p_2 \cdot x_2)}
	\frac{\partial}{\partial x_1^{\mu_1}} \cdots
	\frac{\partial}{\partial x_1^{\mu_\ell}}
	\vev{ \O^{\mu_1 \ldots \mu_\ell}(0) \timeordering{ \phi(x_1) \phi(x_2) }}.
	\nonumber
\end{eqnarray}
Using the explicit form of the 3-point function~\eqref{eq:3pt:position} together with properties of the scalar 3-point integral derived above, it can be shown that this is equivalent to
\begin{equation}
	p_{1\mu_1} \cdots p_{1\mu_\ell} V_\Delta^{\mu_1 \ldots \mu_\ell}(p_1,p_2) 
	= (-i)^\ell \frac{\left( \frac{\Delta - \ell}{2} \right)_\ell
	\left( d - \Delta_\phi - \frac{\Delta + \ell}{2} \right)_\ell}
	{\left( \Delta_\phi - \frac{\Delta + \ell}{2} \right)_\ell} \,
	V_{\Delta + \ell}(p_1,p_2) 
	\label{eq:3pt:scalarcontraction:2}
\end{equation}
where $V_{\Delta + \ell}$ is the scalar integral of Eq.~\eqref{eq:V}, with the scaling dimension of the operator $\O$ shifted by $\ell$.
The equivalence between Eqs.~\eqref{eq:3pt:scalarcontraction:1} and \eqref{eq:3pt:scalarcontraction:2} implies that
\begin{equation}
	C_{\phi\phi\O} =
	\frac{ i^\ell 2^{2d - 2\Delta_\phi - \Delta - \ell + 1} \pi^{d+1}
	\Gamma\left( \Delta_\phi - \frac{d}{2} \right)^2
	\Gamma\left( \frac{\Delta + \ell}{2} - \Delta_\phi + \frac{d}{2} \right)
	\left( \Delta - 1 \right)_\ell}
	{\Gamma\left( \frac{\Delta + \ell}{2} \right)^2
	\Gamma\left( \Delta_\phi - \frac{\Delta - \ell}{2} \right)
	\Gamma\left( \Delta_\phi + \frac{\Delta + \ell}{2} - \frac{d}{2} \right)
	\Gamma\left( \Delta_\phi + \frac{\Delta - \ell}{2} - d + 1 \right)}.
	\label{eq:CphiphiO}
\end{equation}
Note that $C_{\phi\phi\O}$ need only be defined for even $\ell$, as the 3-point function vanishes by symmetry for odd $\ell$. It is therefore a real coefficient, and so is $V_\Delta^{\mu_1 \ldots \mu_\ell}$.
This result completes the computation of the momentum-space 3-point function.

\subsection{Construction of conformal blocks}

With the knowledge of the 2- and 3-point functions, the conformal blocks defined in the introduction can now be read directly from Eq.~\eqref{eq:4pt:factorization}. 
First, using the explicit form of the tensor $\Pi$ given in Eq.~\eqref{eq:Pi}, it can be verified that the 3-point function involving the shadow states~\eqref{eq:shadowstate} is related to the 3-point function constructed with the ordinary state. Explicitly, we find
\begin{equation}
	\Pi^\Delta_{\mu_1 \ldots \mu_\ell, \nu_1 \ldots \nu_\ell}(q) 
	V_\Delta^{\nu_1 \ldots \nu_\ell}(q,r)
	= \frac{1}{C_\O} V_{(d - \Delta)\mu_1 \ldots \mu_\ell}(q,r)
\end{equation}
where $C_\O$ is the constant defined in Eq.~\eqref{eq:CO}.
Therefore, the conformal blocks take the simple form
\begin{equation}
	G_{\Delta, \ell}(x) = 
	\frac{V_\Delta^{\mu_1 \ldots \mu_\ell}(q,r)
	V_{(d - \Delta)\mu_1 \ldots \mu_\ell}(q,r')}{s^{2 \Delta_\phi- 3d/2} C_\O}
	\label{eq:VV}
\end{equation}
where the various momenta are given by $q = p_1 + p_2 = -(p_3 + p_4)$, $r = p_1 - p_2$ and $r' = p_3 - p_4$. This basis of vectors is convenient as all scalar products take a simple form:
\begin{equation}
	-q^2 = r^2 = r'^2 = s,
	\qquad\qquad
	q \cdot r = q \cdot r' = 0,
	\qquad\qquad
	r \cdot r' = - s \, x.
\end{equation}
In particular, $x$ measures the only non-trivial angle between $r$ and $r'$, and all momenta are normalized in units of $s$ so that the dependence on $s$ disappears in Eq.~\eqref{eq:VV}.
In our conventional notation $G_{\Delta,\ell}(x) = \N_{\Delta, \ell} \, g_{\Delta, \ell}(x)$, the constant $\N_{\Delta, \ell}$ is directly related to the normalization of the 2- and 3-point functions by
\begin{equation}
	\N_{\Delta, \ell} = \frac{C_{\phi\phi\O}^2}{C_\O},
\end{equation}
with $C_\O$ and $C_{\phi\phi\O}$ given in Eqs.~\eqref{eq:CO} and \eqref{eq:CphiphiO} respectively.
Evaluating the polynomial $g_{\Delta, \ell}(x)$ is a straightforward exercise of combinatorics, albeit a delicate one due to the presence of the trace terms in Eq.~\eqref{eq:3pt}. We find
\begin{equation}
	g_{\Delta,\ell}(x) = 
	\sum_{n = 0}^{\lfloor \ell / 2 \rfloor} \mathcal{X}_{\ell,n} \, x^{\ell-2n}
	\label{eq:g:repeated}
\end{equation}
with coefficients
\begin{eqnarray}
	\mathcal{X}_{\ell,n} & = & \frac{\ell!}
	{2^{4n} (\ell - 2n)! \Big( \frac{3 - \Delta - \ell}{2} \Big)_n
	\Big( \frac{3 - \widetilde{\Delta} - \ell}{2} \Big)_n}
	\nonumber \\
	&& \quad \times
	\sum_{k = 0}^n \frac{(-1)^k 2^{2k} (2n - 2k)!}{k! \left[ (n-k)! \right]^2}
	\frac{\left( \frac{2 - \Delta - \ell}{2} \right)_k
	\left( \frac{2 - \widetilde{\Delta} - \ell}{2} \right)_k}
	{\left( \frac{d - 2}{2} + \ell - k \right)_k}.
	\label{eq:X:sum}
\end{eqnarray}
This definition is equivalent to the generalized hypergeometric function \eqref{eq:X}.
The first few polynomials are
\begin{eqnarray}
	g_{\Delta, 0}(x) & = & 1,
	\label{eq:g0}
	\\
	g_{\Delta, 2}(x) & = & x^2
	- \frac{\Delta (d - \Delta) - d}
	{d (\Delta - 1) (d - \Delta - 1)},
	\label{eq:g2}
	\\
	g_{\Delta, 4}(x) & = &  x^4
	-6 \, \frac{\Delta (d - \Delta) + d}
	{(d + 4) (\Delta + 1) (d - \Delta + 1)} \, x^2
	\nonumber \\
	&& + 3 \, \frac{\Delta^2 (d - \Delta)^2 - (d + 2)(d - 4)}
	{(d+2) (d+4)(\Delta - 1) (\Delta + 1) (d - \Delta - 1) (d - \Delta + 1)}.
	\label{eq:g4}
\end{eqnarray}
As can be seen, the coefficient of the leading term in $x$ satisfies
\begin{equation}
	\mathcal{X}_{\ell,0} = 1
\end{equation}
for any $\ell$.
The $g_{\Delta,\ell}(x)$ are even polynomials, which realizes the crossing symmetry $t \leftrightarrow u$ of the 4-point function, and they are obviously invariant under the shadow transformation
\begin{equation}
	g_{\Delta, \ell}(x) = g_{d - \Delta, \ell}(x).
	\label{eq:shadow}
\end{equation}
An important property of these polynomials, which is not obvious from their definition, is that they are positive in the forward scattering regime $x = 1$,
\begin{equation}
	g_{\Delta, \ell}(1) \geq 0.
	\label{eq:forwardlimit}
\end{equation}
This is because the conformal block is the norm of a state in that limit~\cite{Gillioz:2016jnn, Gillioz:2018kwh}.

The various special cases listed in the introduction can be straightforwardly obtained from Eq.~\eqref{eq:X:sum}. 
At large $d$, keeping the quantity $\Delta - \ell - d + 2$ fixed, all the terms subleading in $x$ in the polynomial vanish,
\begin{equation}
	\mathcal{X}_{\ell,0} = 1,
	\qquad\qquad
	\mathcal{X}_{\ell,n} \xrightarrow{d \to \infty} 0
	\quad (n > 0).
\end{equation}
This property follows from the simple form of the 3-point function \eqref{eq:3pt} when $\Delta$ is large (also valid at large $\widetilde{\Delta}$, i.e.~when $\Delta \to -\infty$),
\begin{equation}
	V_\Delta^{\mu_1 \ldots \mu_\ell}(q,r)
	\xrightarrow{|\Delta| \to \infty}
	C_{\phi\phi\O} \, s^{\Delta_\phi - d + (\Delta - \ell)/2}
	\left[ r^{\mu_1} \cdots r^{\mu_\ell} - \text{traces} \right]
	\label{eq:3pt:largeDelta}
\end{equation}
and from the fact that all trace terms can be neglected when $d \to \infty$, so that the product \eqref{eq:VV} of the 3-point functions becomes trivial.
At large $\Delta$ but finite $d$, the conformal block is obtained squaring the 3-point function~\eqref{eq:3pt:largeDelta},
which reproduces the Gegenbauer polynomial of Eq.~\eqref{eq:g:largeDelta}.
Alternatively, it can be seen that the sum~\eqref{eq:X:sum} is dominated by the term $k = n$ in that limit, so that
\begin{equation}
	\mathcal{X}_{\ell,n} \xrightarrow{\Delta \to \infty}
	\frac{(-1)^n \ell!}{2^{2n} n! (\ell - 2n)!
	\left( \frac{d-2}{2} + \ell - n \right)_n}.
\end{equation}
Conversely, when $\Delta$ approaches the unitarity bound value $d - 2 + \ell$, only the term $k = 0$ contributes to the sum~\eqref{eq:X:sum}, and one obtains
\begin{equation}
	\mathcal{X}_{\ell,n} \xrightarrow{\Delta \to d - 2 + \ell}
	\frac{(-1)^n \ell!}{2^{2n} n! (\ell - 2n)!
	\left( \frac{d-3}{2} + \ell - n \right)_n},
\end{equation}
corresponding to the other Gegenbauer polynomial~\eqref{eq:g:unitaritybound}.
The fact that these last two limits differ by one unit of spacetime dimension can be understood as follows: When $\Delta$ saturates the unitarity bound, the operator $\O$ is a conserved current, and the states that it defines satisfy therefore $q_{\mu_a} | \O^{\mu_1 \ldots \mu_\ell}(q) \rangle = 0$, for all $a = 1, \ldots, \ell$. In the center-of-mass frame in which $q = (1, 0, \ldots, 0)$, only the states with spatial indices $| \O^{i_1 \ldots i_\ell}(q) \rangle$ are non-null. These states transform as traceless symmetric tensors under the subgroup $\SO(d) \subset \SO(d,1)$. Moreover, the 3-point function projected onto this subspace takes the form of Eq.~\eqref{eq:3pt:largeDelta} up to terms that ensure the conservation property, which explains why one recovers a Gegenbauer polynomial in one less dimension.

The appearance of the Gegenbauer polynomials $\mathcal{C}_\ell^{(d-3)/2}(x)$ is not a surprise, as it establishes a connection with a different expansion of the momentum-space 4-point function, namely the partial wave expansion in which intermediate states are organized in terms of their angular momentum.%
\footnote{We thank Jo\~ao Penedones for pointing this out.}
The connection between the two expansions is not simple, as a single conformal block contains intermediate descendant states with arbitrarily large spin, and conversely a given partial wave receives contribution from a (presumably infinite) tower of primary operators.
Nevertheless, it turns out that the polynomials $g_{\Delta, \ell}(x)$ admit a relatively simple decomposition in terms of partial waves, in the form
\begin{equation}
	g_{\Delta, \ell}(x)
	= \sum_{n = 0}^{\lfloor \ell / 2 \rfloor}
	\frac{(-1)^n (2n)! \ell !}{2^{\ell + 2n + 1} (n!)^2}
	\frac{d - 3 + 2\ell - 4n}
	{\left( 2 - \frac{d}{2} - \ell \right)_n
	\left( \frac{d-3}{2} \right)_{\ell - n +1}}
	\frac{\Big( \frac{2 - \Delta - \ell}{2} \Big)_n
	\Big( \frac{2 - \widetilde{\Delta} - \ell}{2} \Big)_n}
	{\Big( \frac{3 - \Delta - \ell}{2} \Big)_n
	\Big( \frac{3 - \widetilde{\Delta} - \ell}{2} \Big)_n} \,
	\mathcal{C}_{\ell - 2n}^{(d-3)/2}(x).
\end{equation}
The coefficients relating the $g_{\Delta, \ell}(x)$ to the Gegenbauer polynomials $\mathcal{C}_n^{(d-3)/2}(x)$ are rational functions of the scaling dimension $\Delta$.
This is not the case if one tries to expand the $g_{\Delta, \ell}(x)$ in terms of a different basis of polynomials, as for instance the $\mathcal{C}_n^{(d-2)/2}(x)$.

The large and small $\Delta$ limits of the conformal blocks are particularly interesting due to the orthogonality of the Gegenbauer polynomials. In the next section we exploit this property to invert the OPE in a case where only conserved currents appear.


\section{An application: OPE inversion in the free scalar theory}
\label{sec:freefields}

Free theories are interesting from an algebraic CFT point-of-view:
for instance in the theory of a free scalar field $\phi(x)$, there are infinitely many primary operators entering the $\phi \times \phi$ OPE. We can write them schematically as the normal-ordered product of two fields with derivatives acting on either of them,
\begin{equation}
	\O^{\mu_1 \ldots \mu_\ell}(x)
	\sim ~ : \phi(x) \partial^{\mu_1} \cdots \partial^{\mu_\ell} \phi(x) :.
	\label{eq:doubletraceops}
\end{equation}
An explicit construction shows that there is exactly one such operator for every even spin $\ell$, and none for odd $\ell$, in accordance with the fact that the equation of motion $\square \phi = 0$ forbids the contraction of indices. With the exception of the scalar operator ($\ell = 0$), all these operator are higher-spin conserved currents, as their scaling dimension saturates the unitarity bound, $\Delta_{\O} = d - 2 + \ell$.
The existence of these double-trace operators is needed to reproduce the Gaussian nature of the 4-point function in terms of ordinary conformal blocks.

In momentum space, the correlator of 4 free fields is trivial: $\M$ is a sum of delta functions and it does not have an imaginary part. This is consistent with the vanishing of the coefficient $\N_{\Delta, \ell}$, visible in Eq.~\eqref{eq:N}. The free scalar field theory contains however other scalar operators whose correlators are not Gaussian, and for which the momentum-space conformal blocks are interesting: this is for instance the case of the first operator with $\ell = 0$ in Eq.~\eqref{eq:doubletraceops}, namely
\begin{equation}
	\O(x) = \frac{1}{\sqrt{2}} : \phi(x)^2 :
	\label{eq:phisquarednormalization}
\end{equation}
where the numerical factor is fixed by the standard normalization condition~\eqref{eq:scalar2pt:position:Feynman} of the 2-point function. In this section, we will discuss the conformal block expansion of the 4-point function of $\O(x)$, and show how the results of the previous sections can be used to compute OPE coefficients.

\begin{figure}
	\centering
	\begin{tabular}{@{}c@{}c@{}c@{}}
	\begin{fmffile}{boxdiagram1}
		\begin{fmfgraph*}(130,40)
			\fmfset{arrow_len}{3mm}
			\fmfleft{p2,p1}
			\fmfright{p4,p3}
			\fmf{scalar,label=$p_1$,l.s=left}{p1,v1}
			\fmf{scalar,label=$p_2$,l.s=right}{p2,v2}
			\fmf{scalar,label=$p_4$,l.s=right}{p3,v3}
			\fmf{scalar,label=$p_3$,l.s=left}{p4,v4}
			\fmf{plain}{v1,v3}
			\fmf{plain}{v4,v2}
			\fmffreeze
			\fmf{plain}{v2,v1}
			\fmf{plain}{v3,v4}
			\fmfdot{v1,v2,v3,v4}
		\end{fmfgraph*}
	\end{fmffile}
	&
	\begin{fmffile}{boxdiagram2}
		\begin{fmfgraph*}(130,40)
			\fmfset{arrow_len}{3mm}
			\fmfleft{p2,p1}
			\fmfright{p4,p3}
			\fmf{scalar,label=$p_1$,l.s=left}{p1,v1}
			\fmf{scalar,label=$p_2$,l.s=right}{p2,v2}
			\fmf{scalar,label=$p_4$,l.s=right}{p3,v3}
			\fmf{scalar,label=$p_3$,l.s=left}{p4,v4}
			\fmf{phantom}{v1,v3}
			\fmf{phantom}{v4,v2}
			\fmffreeze
			\fmf{plain}{v1,v3,v2,v4,v1}
			\fmfdot{v1,v2,v3,v4}
		\end{fmfgraph*}
	\end{fmffile}
	&
	\begin{fmffile}{boxdiagram3}
		\begin{fmfgraph*}(130,40)
			\fmfset{arrow_len}{3mm}
			\fmfleft{p2,p1}
			\fmfright{p4,p3}
			\fmf{scalar,label=$p_1$,l.s=left}{p1,v1}
			\fmf{scalar,label=$p_2$,l.s=right}{p2,v2}
			\fmf{scalar,label=$p_4$,l.s=right}{p3,v3}
			\fmf{scalar,label=$p_3$,l.s=left}{p4,v4}
			\fmf{phantom}{v1,v3}
			\fmf{phantom}{v4,v2}
			\fmffreeze
			\fmf{plain}{v1,v4,v3,v2,v1}
			\fmfdot{v1,v2,v3,v4}
		\end{fmfgraph*}
	\end{fmffile}
	\\[1.5em]
	$\M_s$ & $\M_t$ & $\M_u$
	\end{tabular}
	\caption{The 3 connected Feynman diagrams that contribute
	to the 4-point function of the operator $\O \sim \phi^2$ in the free scalar theory.
	The dashed lines indicate the external operator $\O$,
	while the solid lines represent propagators of the free field $\phi$.}
	\label{fig:boxdiagrams}
\end{figure}
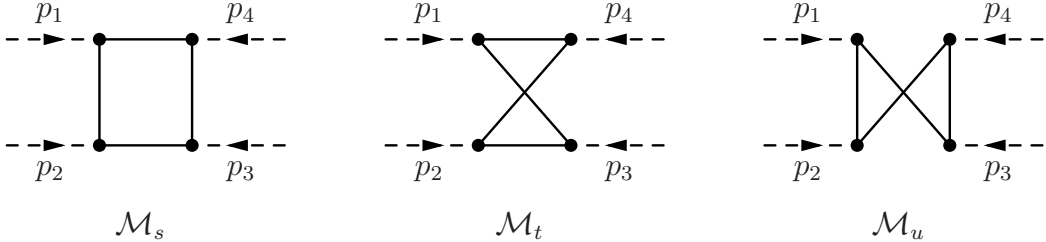
In the free theory, the 4-point function can be computed explicitly in terms of Feynman diagrams.
The three connected diagrams that enter the computation are shown in Fig.~\ref{fig:boxdiagrams}, and $\M$ is given by their sum
\begin{equation}
	\M = \M_s + \M_t + \M_u.
\end{equation}
The computation of each individual diagram is described in details in Appendix~\ref{sec:loopintegral}. It should be noted that the loop integrals are UV divergent in spacetime dimension $d \geq 8$, and IR divergent in $d \leq 6$. The UV divergence arises from the fact that the source for the operator $\O$ has dimension 2, and that it possible in $d = 8$ (and in even $d > 8$) to write a counterterm involving 4 sources in the action. This counterterm is nevertheless real, and the imaginary part of the 4-point function must therefore be finite in all $d > 6$. 
The explicit computation of Appendix~\ref{sec:loopintegral} yields the result
\begin{equation}
	G(x) = 
	\frac{2^{11-d} \pi^{3d/2+1}}
	{(d-4) \Gamma(d-3) \Gamma\left( \frac{d-2}{2} \right)^3}
	\Big[ \,_2F_1\left( 1, 1; \tfrac{d-2}{2}; \tfrac{1+x}{2} \right)
	+ \,_2F_1\left( 1, 1; \tfrac{d-2}{2}; \tfrac{1-x}{2} \right) \Big],
	\label{eq:G:freescalar}
\end{equation}
which is indeed finite in $d > 6$. The hypergeometric $_2F_1$ functions take a simple form in all integer dimensions, for instance in $d = 5$,
\begin{equation}
	G(x) \Big|_{d = 5} = \frac{512 \pi^8}{\sqrt{1 - x^2}}.
	\label{eq:G:freescalar:5d}
\end{equation}
In this case, the IR divergence only shows up at $x = \pm 1$. The same observation can be made in $d = 6$. For generic values of $x$, $G(x)$ can actually be analytically continued in $d$ from $d > 6$ down to $d > 4$. In $d = 4$, the imaginary part of $\M$ diverges for all $x$.

The conformal block expansion derived in this work can now be applied to $G(x)$.
There are two types of primary operators that enter the $\O \times \O$ OPE: an infinite series of the form $\partial^n \phi^4$, and the operators~\eqref{eq:doubletraceops} of the form $\partial^n \phi^2$. The former do not contribute to the imaginary part of the 4-point function of $\O$, as they have double-trace dimensions for which the momentum-space blocks vanish. On the other hand, the operators $\O^{\mu_1 \ldots \mu_\ell}$ give non-vanishing contributions.
Using the fact that they have scaling dimension $\Delta = d - 2 + \ell$, the conformal block expansion takes the form
\begin{equation}
	G(x) = \sum_{\ell = 0}^\infty \lambda_\ell^2 \,
	\frac{2^{d + \ell + 4} \pi^{3d/2} (l!)^3
	\Gamma\left( \frac{d-3}{2} \right) 
	\Gamma\left( \frac{d-1}{2} + \ell \right)}
	{(d-4)^2 \Gamma\left( \frac{d-2}{2} + \ell \right)
	\Gamma\left( d - 3 + \ell \right)^3} \,
	\mathcal{C}_\ell^{(d-3)/2}(x),
\end{equation}
where $\lambda_\ell$ indicates the OPE coefficient between two scalars $\O$ of Eq.~\eqref{eq:phisquarednormalization} and one spin-$\ell$ operator of Eq.~\eqref{eq:doubletraceops}.
The orthogonality of Gegenbauer polynomials can then be used to write an inversion formula in the form of an integral of $G(x)$ over the interval $x \in [-1, 1]$, namely
\begin{equation}
	\lambda_\ell^2 = \frac{(d-4)^2 \Gamma\left( \frac{d-2}{2} + \ell \right)
	\Gamma\left( d - 3 + \ell \right)^2}
	{2^{\ell + 8} \pi^{3d/2+1} (l!)^2
	\left( \frac{d-3}{2} \right)_\ell}
	\int_{-1}^1 dx \left( 1 - x^2 \right)^{(d-4)/2} 
	\mathcal{C}_\ell^{(d-3)/2}(x)
	G(x).
	\label{eq:inversion}
\end{equation}
Plugging in the expression \eqref{eq:G:freescalar} for $G(x)$, one finds
\begin{equation}
	\lambda_\ell^2 = \big[ 1 + (-1)^\ell \big]
	\frac{2^\ell \left( \frac{d-2}{2} \right)_\ell^2}
	{\ell! \, \left( d + \ell - 3 \right)_\ell}.
\end{equation}
These OPE coefficients are found to be in agreement with previous computations~\cite{Fitzpatrick:2011dm, Fitzpatrick:2012yx, Gillioz:2016jnn}. It should be noted that they are regular in any dimension, including $d = 3$ and 4, as a consequence of the analyticity in $d$ of our method.

\begin{figure}
	\centering
	\includegraphics[width=0.48\linewidth]{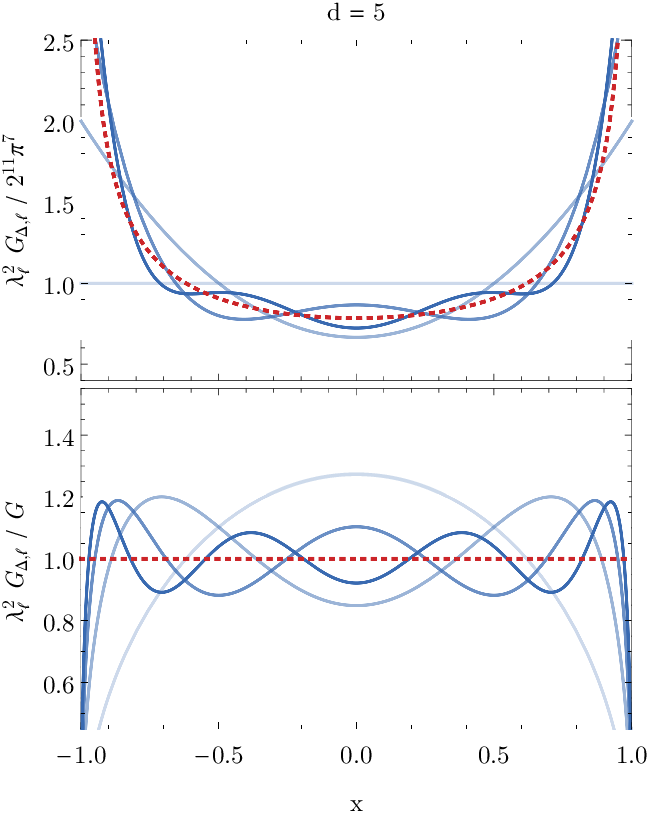}
	\hfill
	\includegraphics[width=0.48\linewidth]{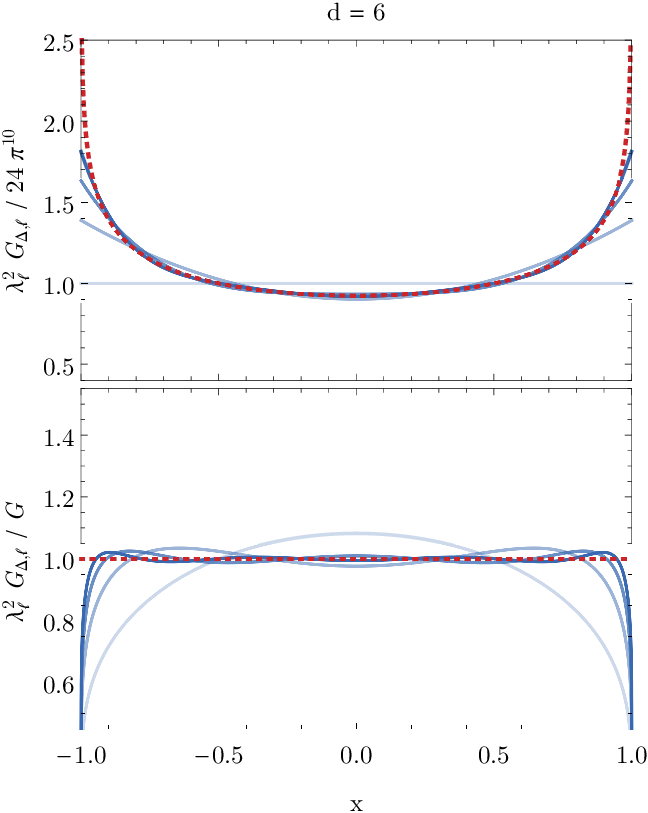}
	\caption{The combined contribution of conformal blocks
	up to a maximal spin $\ell_\text{max} = 0, 2, 4, 6$ (from lighter to darker blue lines)
	to the imaginary part of $\M$ for the operator $\phi^2$ in the free scalar theory,
	in $d = 5$ (left) and $d = 6$ (right) spacetime dimensions.
	The red dotted line indicates the full result given in Eq.~\eqref{eq:G:freescalar}.
	The upper panels correspond to the actual value of the conformal blocks,
	while the lower panels show their relative contribution.}
	\label{fig:freetheories}
\end{figure}%
Finally, we illustrate in Fig.~\ref{fig:freetheories} the convergence of the momentum-space OPE in this free theory example, showing the combined contribution of operators with spin 0, 2, 4 and 6 to $G(x)$ in $d = 5$ and 6 dimensions.
The convergence is clearly fast in $d = 6$, and this property carries on in $d > 6$. It slows down when $d$ approaches the critical dimension $d = 4$, where every single conformal block diverges individually, and so does $G(x)$.
Below that critical dimension, in $d = 3$, the imaginary part of $\M$ computed from the Feynman diagram result~\eqref{eq:G:freescalar:5d} is finite. The individual conformal blocks are also finite, but it can be verified that they grow with the spin of the intermediate operator. The OPE does not converge in this case.


\section{Conclusions}
\label{sec:conclusions}

In this work, we have computed conformal blocks for the momentum-space 4-point function of identical scalar operators in the light-cone limit.
More than the result itself, the main message that we would like to carry is the simplicity of the method: using translation invariance in the form of momentum conservation, together with a particularly simple realization of the shadow operator formalism, the conformal blocks can be obtained by direct multiplication of 3-point functions.
The result is a polynomial in the cosine of the scattering angle, with coefficients given in a closed-form expression valid in any spacetime dimension $d$.
This direct computation method is expected to stay relatively simple for conformal blocks of external operators carrying spin, even though it would be interesting to have an alternative formulation of the momentum-space conformal blocks, either as the solution a differential equation~\cite{Dolan:2003hv}, or possibly using recursion relations~\cite{Kos:2013tga, Hogervorst:2013sma, Kos:2014bka, Penedones:2015aga, Costa:2016xah}. 

There are several features of the momentum-space blocks that could have interesting applications in conformal field theory.
The positivity of the blocks at $x = \pm 1$ has already been exploited in Refs.~\cite{Gillioz:2016jnn, Gillioz:2018kwh} to derive positive sum rules for anomaly coefficients.
Their orthogonality for low and high scaling dimensions of the intermediate operator is suggestive of more general orthogonality properties, which should be studied in relation with OPE inversion formulae~\cite{Caron-Huot:2017vep, Simmons-Duffin:2017nub, Kravchuk:2018htv, Alday:2016njk, Cardona:2018nnk, Cardona:2018dov}.
But more importantly, the primary use of conformal blocks could be in a momentum-space formulation of the bootstrap program. It should be noted however that while crossing symmetry in the channel $t \leftrightarrow u$ is automatically realized through the parity property of the polynomials $g_{\Delta, \ell}(x)$, there is no obvious crossing equation for the channels $s \leftrightarrow t$ and  $s \leftrightarrow u$, as they relate the imaginary part of $\M$ to its real part, for which there is no conformal block expansion. 
A possible solution to this problem would be to exploit the analyticity properties of the 4-point function.
We leave the study of these questions for future work.

\begin{table}
	\centering
	\begin{tabular}{|@{\quad}c@{\qquad}l@{\qquad}c@{\quad}|
	c@{\qquad}r@{\qquad}cc@{\qquad}r@{\qquad}c|}
		\hline
		\multirow{2}{*}{$\ell$} &
		\multirow{2}{*}{~~$\Delta$} & 
		\multirow{2}{*}{$\lambda_{\sigma\sigma\O}$} &
		\multicolumn{3}{|c}{$\lambda_{\sigma\sigma\O}^2 G_{\Delta,\ell}(0)$} &
		\multicolumn{3}{c|}{$\lambda_{\sigma\sigma\O}^2 G_{\Delta,\ell}(1)$} \\
		&&& \multicolumn{3}{|c}{$\cdot 10^{-10}$}
		& \multicolumn{3}{c|}{$\cdot 10^{-10}$} \\
		\hline
		\hline
		0 & 1.412625 & 1.051854 && 0.0 &&& 0.0 & \\
		0 & 3.82968 & 0.053012 && 5.5 &&& 5.5 & \\
		0 & 6.8956 & 0.000734 && 9.9 &&& 9.9 & \\
		0 & 7.2535 & 0.000162 && 3.2 &&& 3.2 & \\
		\hline
		2 & 3 & 0.652276 && 0.0 &&& 0.0 & \\
		2 & 5.50915 & 0.021149 && $-19.3$ &&& 35.2 & \\
		2 & 7.0758 & 0.000955 && $-0.3$ &&& 0.6 & \\
		\hline
		4 & 5.022665 & 0.276304 && 0.0 &&& 0.0 & \\
		4 & 6.42065 & 0.007821 && 6.0 &&& 11.8 & \\
		4 & 7.38568 & 0.009510 && 33.4 &&& 74.1 & \\
		\hline
		6 & 7.028488 & 0.125933 && 0.0 &&& 0.0 & \\
		\hline
	\end{tabular}
	\caption{List of operators with dimension $\Delta \leq 8$ that enter the $\sigma \times \sigma$ OPE in the
	critical 3d Ising model, with their spin $\ell$, scaling dimension $\Delta$,
	OPE coefficient $\lambda_{\sigma\sigma\O}$,
	and the contribution of these operator to the imaginary part of the 4-point function  at $x = 0$ and $x = 1$.
	$\sigma$ is the lowest-dimension scalar operator, with $\Delta_\sigma = 0.5181489$.
	The data on the left-hand side of this table is taken from Ref.~\cite{Simmons-Duffin:2016wlq},
	while the right-hand side is computed from the conformal blocks.
	}
	\label{tab:Isingdata}
\end{table}%
Finally, in spite of the interesting features described above, there is an important downside to the use of momentum-space conformal blocks that must be mentioned:
it is unclear in which situations the use of the completeness relation \eqref{eq:completenessrelation:shadow} leads to a convergent series expansion. 
The free scalar field theory setup of Section~\ref{sec:freefields} provides a concrete example of this problem in $d = 3$: even though each conformal block is finite, their sum does not converge.
It is understood in this case how the the divergence is related to the IR singularities of a loop integral.
We do not know however how to address the problem of possible IR divergences in interacting theories.
The simplest case that we can examine is the Ising model in $d = 3$. Since the spectrum of operators and the OPE coefficients are known for low-dimension operators entering the OPE of the lightest scalar~\cite{Simmons-Duffin:2016wlq}, we can evaluate the first few conformal blocks and check if they hint towards a convergent series. The values of $G_{\Delta, \ell}$ in the forward limit $x = 1$ and in the right-angle scattering case $x = 0$ for each operator are given in Table~\ref{tab:Isingdata}. The inspection of this data is however inconclusive. For each spin, the leading operator gives a very small contribution to the 4-point function, as its scaling dimension is very close to the double-trace  limit $\Delta \approx 2 \Delta_\sigma + \ell + 2n$. Among the remaining operators, there is no clear hierarchy that could indicate a convergent expansion, although operators of low twist seem to give overall larger contributions to the 4-point function.
The question of the OPE convergence in momentum space will have to remain open for now.


\acknowledgments
This work has been supported by the Swiss National Science Foundation through the NCCR SwissMAP.


\appendix
\section{One-loop integral in the free scalar field theory}
\label{sec:loopintegral}

In this appendix, we briefly outline the computation of the one-loop Feynman diagrams that appear in Section~\ref{sec:freefields}.
The diagrams under consideration are a special case of the usual scalar box integral \cite{tHooft:1978jhc}, in which all internal and external propagators are massless. This is a well-known integral, but it usually not considered in the massless limit due to infrared divergences in $d = 4$. We will therefore detail its evaluation here.

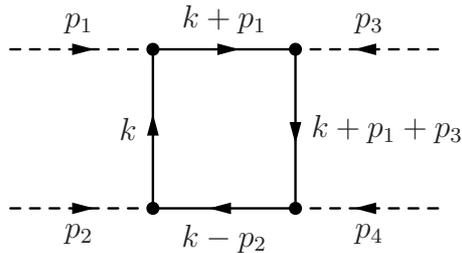
\begin{figure}
	\centering
	\begin{fmffile}{boxdiagram1L}
		\fmfframe(20,20)(20,20){
		\begin{fmfgraph*}(200,60)
			\fmfset{arrow_len}{3mm}
			\fmfleft{p2,p1}
			\fmfright{p4,p3}
			\fmf{scalar,label=$p_1$,l.s=left}{p1,v1}
			\fmf{scalar,label=$p_2$,l.s=right}{p2,v2}
			\fmf{scalar,label=$p_3$,l.s=right}{p3,v3}
			\fmf{scalar,label=$p_4$,l.s=left}{p4,v4}
			\fmf{fermion,label=$k + p_1$,l.s=left}{v1,v3}
			\fmf{fermion,label=$k - p_2$,l.s=left}{v4,v2}
			\fmffreeze
			\fmf{fermion,label=$k$,l.s=left}{v2,v1}
			\fmf{fermion,label=$k + p_1 + p_3$,l.s=left}{v3,v4}
			\fmfdot{v1,v2,v3,v4}
		\end{fmfgraph*}}
	\end{fmffile}
	\caption{Reproduction of the diagram $\M_u$ of Fig.~\ref{fig:boxdiagrams} with labels and arrows
	indicating our choice for the loop momenta.}
	\label{fig:boxdiagramwithlabels}
\end{figure}%
We begin with the Feynman diagram of Fig.~\ref{fig:boxdiagramwithlabels}. The other two diagrams are related to this one by crossing.
Since we use the standard CFT normalization~\eqref{eq:scalar2pt:position:Feynman} of the 2-point function in position space, the propagator in momentum space comes with an additional normalization factor compared to usual Feynman rules, which can be read off directly from Eq.~\eqref{eq:scalar2pt:Feynman} setting the scaling dimension $\Delta$ to its free field value,
\begin{equation}
	F_{(d-2)/2}(p) = \frac{4 \pi^{d/2}}{\Gamma\left( \frac{d-2}{2} \right)}
	\frac{i}{-p^2 + i \epsilon}.
\end{equation}
The integral can therefore be written as
\begin{equation}
	\M_u = \frac{2^{10} \pi^{2d}}{\Gamma\left( \frac{d-2}{2} \right)^4}
	\int \frac{d^dk}{(2\pi)^d}
	\frac{1}{k^2 (k + p_1)^2 (k + p_1 + p_2)^2 (k - p_2)^2}.
	\label{eq:Feynmanintegral}
\end{equation}
This expression includes a factor of $4$ coming from the normalization~\eqref{eq:phisquarednormalization} of the operator $\phi^2$.
Introducing Feynman parameters and shifting the integration momentum appropriately, the integral can be rewritten as
\begin{equation}
	\M_u = \frac{3 \, 2^9 \pi^{2d}}{\Gamma\left( \frac{d-2}{2} \right)^4}
	\left( \prod_{i=1}^4 \int_0^1 d\lambda_i \right)
	\delta(\lambda_1 + \lambda_2 + \lambda_3 + \lambda_4 - 1)
	\int \frac{d^dk}{(2\pi)^d} \frac{1}{(k^2 - \lambda_1 \lambda_2 s - \lambda_3 \lambda_4 t - i \epsilon)^4}.
\end{equation}
Note that we have introduced the $i \epsilon$ prescription that was implicit in Eq.~\eqref{eq:Feynmanintegral}.
Evaluating the momentum integral, we obtain
\begin{eqnarray}
	\M_u & = & \frac{2^{8-d} \pi^{3d/2} \Gamma\left( \frac{8-d}{2} \right)}
	{\Gamma\left( \frac{d-2}{2} \right)^4}
	\\
	&& \times \int_0^1 d\lambda_1 \int_0^{1-\lambda_1} d\lambda_2 \int_0^{1-\lambda_1-\lambda_2} d\lambda_3
	\Big[ -\lambda_1 \lambda_2 s - \lambda_3 (1-\lambda_1 - \lambda_2 - \lambda_3) t - i \epsilon \Big]^{(d-8)/2}.
	\nonumber
\end{eqnarray}
The gamma function in the numerator is a reminder of the fact that the integral is UV divergent in even $d \geq 8$, which is obvious in Eq.~\eqref{eq:Feynmanintegral}.
After the change of variables defined by
$\lambda_1 = (1-\omega_1)(1-\omega_3)$,
$\lambda_2 = \omega_2 \omega_3$
and $\lambda_3 = \omega_1 (1-\omega_3)$,
this becomes
\begin{eqnarray}
	\M_u & = & \frac{2^{8-d} \pi^{3d/2} \Gamma\left( \frac{8-d}{2} \right)}
	{\Gamma\left( \frac{d-2}{2} \right)^4}
	\\
	&& \times \int_0^1 d\omega_1 d\omega_2 d\omega_3 \big[ \omega_3 (1-\omega_3) \big]^{(d-6)/2}
	\Big[ -(1 - \omega_1) \omega_2 s - \omega_1 (1 - \omega_2) t - i \epsilon \Big]^{(d-8)/2}.
	\nonumber
\end{eqnarray}
The integral over $\omega_3$ factorizes and can be evaluated explicitly in $d > 4$ to get
\begin{equation}
	\M_u = \frac{2^{10-d} \pi^{3d/2} \Gamma\left( \frac{8-d}{2} \right)}
	{(d-4) \Gamma(d-3) \Gamma\left( \frac{d-2}{2} \right)^2}
	\int_0^1 d\omega_1 d\omega_2
	\Big[ -(1 - \omega_1) \omega_2 s - \omega_1 (1 - \omega_2) t - i \epsilon \Big]^{(d-8)/2}.
\end{equation}
Next, note that the term in square brackets changes sign over the region of integration, since $s > 0$ and $t < 0$ in the kinematical configuration that we consider. We split therefore the integral over $\omega_2$ into two regions $0 \leq \omega_2 \leq \omega_2^*$ and $\omega_2^* \leq \omega_2 \leq 1$, where $\omega_2^* = -\omega_1 t / (s + \omega_1 u) \in [0,1]$, and rescale in each case the interval with a change of variable $\omega_2 \to \omega_2^* \omega_2$ and $\omega_2 \to \omega_2^* - (1 - \omega_2^*) \omega_2$ respectively, after which we obtain
\begin{eqnarray}
	\M_u & = & \frac{2^{10-d} \pi^{3d/2} \Gamma\left( \frac{8-d}{2} \right)}
	{(d-4) \Gamma(d-3) \Gamma\left( \frac{d-2}{2} \right)^2}
	\nonumber \\
	&& \times \left[ (-s - i \epsilon)^{(d-8)/2} \int_0^1 d\omega_1 d\omega_2
	\frac{\omega_1^{(d-6)/2} \omega_2^{(d-8)/2}}{1 + (1 - \omega_1) u/s}
	+ (s \leftrightarrow t) \right].
\end{eqnarray}
The two regions are related by crossing $s \leftrightarrow t$, in agreement with the symmetries of the diagram in Fig.~\ref{fig:boxdiagramwithlabels}. The integral over $\omega_2$ is now trivial, but it is only convergent in $d > 6$: in $d \leq 6$, $\M_u$ has an IR divergence.
The remaining integral over $\omega_1$ corresponds to a hypergeometric $_2F_1$ function, and we get
\begin{eqnarray}
	\M_u & = & \frac{2^{12-d} \pi^{3d/2} \Gamma\left( \frac{8-d}{2} \right)}
	{(d-4)^2 (d-6) \Gamma(d-3) \Gamma\left( \frac{d-2}{2} \right)^2}
	\label{eq:Mu:appendix}
	\\
	&& \times \left[ (-s - i \epsilon)^{(d-8)/2} \,_2F_1\left( 1, 1; \tfrac{d-2}{2}; - \tfrac{u}{s} \right)
	+ (-t)^{(d-8)/2} \,_2F_1\left( 1, 1; \tfrac{d-2}{2}; - \tfrac{u}{t} \right) \right].
	\nonumber
\end{eqnarray}
This integral is analytic at small $u$, and the hypergeometric functions are real for all physical momenta. Of the two terms in the square brackets, only the first one has an imaginary part, coming from  $(-s)^{(d-8)/2}$. Explicitly, using the notation of Eq.~\eqref{eq:xdef}, we have
\begin{equation}
	\im \M_u =  s^{(d-8)/2}
	\frac{2^{10-d} \pi^{3d/2+1}}
	{(d-4) \Gamma(d-3) \Gamma\left( \frac{d-2}{2} \right)^3}
	\,_2F_1\left( 1, 1; \tfrac{d-2}{2}; \tfrac{1+x}{2} \right).
	\label{eq:imMu:appendix}
\end{equation}
The imaginary part of the box diagram is therefore finite in all $d > 4$ for generic $x$. The forward limit $t \to 0$ (or equivalently $x \to 1$) is divergent for $4 < d \leq 6$, while in $d > 6$ the integral is finite for all scattering angles.

The other two diagrams in Fig.~\ref{fig:boxdiagrams} can be evaluated in a similar fashion. For $\M_t$, the result is simply given by exchanging $t$ and $u$ in Eq.~\eqref{eq:Mu:appendix},
and its imaginary part is given by
\begin{equation}
	\im \M_t =  s^{(d-8)/2}
	\frac{2^{10-d} \pi^{3d/2+1}}
	{(d-4) \Gamma(d-3) \Gamma\left( \frac{d-2}{2} \right)^3}
	\,_2F_1\left( 1, 1; \tfrac{d-2}{2}; \tfrac{1-x}{2} \right).
	\label{eq:imMt:appendix}
\end{equation}
The evaluation of $\M_s$ follows a different path, since the integrand is real all along. For all our purposes, it is therefore sufficient to notice that its imaginary part vanishes,
\begin{equation}
	\im \M_s = 0.
	\label{eq:imMs:appendix}
\end{equation}


\newpage

\bibliography{Bibliography}{}
\bibliographystyle{JHEP}

\end{document}